\documentclass[aps,nofootinbib]{revtex4}

\usepackage{graphicx}

\graphicspath{{FZDimages/}}

\usepackage{bm}
\usepackage{mathbbol}
\usepackage{amsmath,amssymb}
\usepackage{mathrsfs}
\usepackage{slashed}
\usepackage{stmaryrd}
\usepackage{wasysym}

\newcommand{\be}{\begin{equation}}
\newcommand{\ee}{\end{equation}}
\newcommand{\bea}{\begin{eqnarray}}
\newcommand{\eea}{\end{eqnarray}}
\newcommand{\bml}{\begin{mathletters} \baselineskip 10pt}
\newcommand{\eml}{\baselineskip 12pt \end{mathletters}}

\newcommand{\ud}{\mathrm{d}}

\newcommand{\bra}[1]{\langle\,#1\,|}          
\newcommand{\ket}[1]{|\,#1\,\rangle}          

\newcommand{\simleq}{\scriptstyle{\stackrel{<}{\sim}}}

\def\lambdabar{\protect\@lambdabar}
\def\@lambdabar{%
\relax \bgroup
\def\@tempa{\hbox{\raise.73\ht0
\hbox to0pt{\kern.2\wd0\vrule width.7\wd0
height.1pt depth.1pt\hss}\box0}}%
\mathchoice{\setbox0\hbox{$\displaystyle\lambda$}\@tempa}%
{\setbox0\hbox{$\textstyle\lambda$}\@tempa}%
{\setbox0\hbox{$\scriptstyle\lambda$}\@tempa}%
{\setbox0\hbox{$\scriptscriptstyle\lambda$}\@tempa}%
\egroup }

\newcommand{\pad}[2]{\frac{\partial #1}{\partial #2}}

\newcommand{\vc}[1]{\mbox{\boldmath$#1$}}

\newcommand{\ssvc}[1]{\mbox{\scriptsize\boldmath$#1$}}

\newcommand{\Ep}{E_{\ssvc{p}}}

\newcommand{\e}{\mbox{e}}

\begin{document}

\title{Signatures of High-Intensity Compton Scattering}

\author{Chris Harvey}\email{christopher.harvey@plymouth.ac.uk}
\author{Thomas Heinzl}\email{theinzl@plymouth.ac.uk}
\affiliation{School of Mathematics and Statistics, University of Plymouth\\
Drake Circus, Plymouth PL4 8AA, UK}
\author{Anton Ilderton}\email{antoni@maths.tcd.ie}
\affiliation{School of Mathematics, Trinity College \\ Dublin 2, Ireland}

\date{\today}

\begin{abstract}
We review known and discuss new signatures of high-intensity Compton scattering assuming a scenario where a high-power laser is brought into collision with an electron beam. At high intensities one expects to see a substantial red-shift of the usual kinematic Compton edge of the photon spectrum caused by the large, intensity dependent, effective mass of the electrons within the laser beam. Emission rates acquire their global maximum at this edge while neighbouring smaller peaks signal higher harmonics.  In addition, we find that the notion of the centre-of-mass frame for a given harmonic becomes intensity dependent. Tuning the intensity then effectively amounts to changing the frame of reference, going continuously from inverse to ordinary Compton scattering with the centre-of-mass kinematics defining the transition point between the two.
\end{abstract}



\maketitle


\tableofcontents

\section{\label{sec:1}Introduction}

The technological breakthrough of laser chirped-pulse amplification \cite{strickland:1985} has led to unprecedented laser powers and intensities, the current records being about 1 Petawatt (PW) and $10^{22}$ W/cm$^2$, respectively. Within the next few years these are expected to be superseded by an increase of about an order of magnitude each, for instance at the upgraded Vulcan laser facility \cite{Vulcan10PW:2009}. Up to three orders of magnitude may be gained at the planned `Extreme Light Infrastructure' facility \cite{ELI:2009}.  This progress calls for a reassessment of intensity effects in QED and the new prospects of measuring them (see e.g.\ \cite{Heinzl:2008wh}, \cite{Heinzl:2008an} and \cite{Marklund:2008gj} for discussions of strong-field physics at Vulcan and ELI). There is a plethora of strong-field QED processes, which may be roughly categorised into two classes; loop and tree-level processes. The former include strong-field vacuum polarisation, the real part of which describes vacuum birefringence \cite{toll:1952} (for a recent discussion see \cite{Heinzl:2006xc}) while its imaginary part signals Breit-Wheeler pair production \cite{Breit:1934}. Summing all orders of these one-loop diagrams (in the low-energy limit) one obtains the Heisenberg-Euler effective Lagrangian \cite{heisenberg:1936} which in turn yields Schwinger's nonperturbative mechanism of spontaneous pair production from the vacuum \cite{Schwinger:1951nm}. The optical theorem and crossing symmetry relate these one-loop diagrams to tree-level processes such as perturbative pair production, pair annihilation and Compton scattering.

It is well known that one-loop processes are of order $\hbar$ and thus of a genuine quantum nature, while tree level processes generically do have a classical limit. As a result, one can introduce two distinct parameters which characterise the different physics involved. The first parameter is the QED electrical field,
\be \label{ECRIT}
  E_c \equiv \frac{m^2 c^3}{e \hbar}  = 1.3 \times 10^{18} \, \mbox{V/m} \; ,
\ee
first introduced by Sauter \cite{Sauter:1931zz} in his analysis of Klein's paradox \cite{Klein:1929b}. The presence of Planck's constant, $\hbar$, and the speed of light, $c$, show that $E_c$ originates from a relativistic quantum field theory. In an electric field of strength $E_c$ an electron acquires an electromagnetic energy equal to its rest mass  $mc^2$ upon traversing a distance of a Compton wavelength, $\lambdabar_e = \hbar/mc$. Hence, $E_c$ may be viewed as the critical field strength above which vacuum pair production becomes abundant. This is also borne out by Schwinger's pair creation probability given by the tunnelling factor $p \sim \exp (-\pi E_c/E)$ \cite{Schwinger:1951nm} where $E$ denotes the `ambient' electric field one succeeds in achieving. Currently, this is $E \simeq 10^{14}$ V/m implying a huge exponential suppression. The perturbative variant of the Schwinger process, i.e.\ the (strong-field) Breit-Wheeler process \cite{Breit:1934,Reiss:1962} was observed about a decade ago in the SLAC E--144 experiment \cite{Burke:1997ew,Bamber:1999zt}. There a Compton backscattered photon pulse of about 30 GeV was brought into collision with the 50 GeV SLAC electron beam. The huge gamma factor ($\gamma \simeq 10^5$) led to an effective electric field close to the critical one, $E' = \gamma E \simeq E_c$, as seen by the electron in its rest frame.

In this context a second parameter comes into play, the `dimensionless laser amplitude', given as the ratio of the electromagnetic energy gained by an electron across a laser wavelength $\lambdabar$ to its rest mass,
\be \label{A0}
  a_0 \equiv \frac{eE\lambdabar}{mc^2}  \; .
\ee
This is a purely classical ratio which exceeds unity once the electron's quiver motion in the laser beam has become relativistic. It may be generalised to an explicitly Lorentz and gauge invariant expression \cite{Heinzl:2008rh}.  For our present purposes it is sufficient to adopt a useful rule-of-thumb formula expressing $a_0$ in terms of laser power \cite{McDonald:1986zz},
\be
  a_0^2 \simeq 5 \times 10^3 P/\mbox{PW} \; ,
\ee
so that $a_0$ is of order $10^2$ for a laser in the Petawatt class.  SLAC E--144, on the other hand, had $a_0$ of order one, hence by modern standards was in the low-intensity, high-energy regime. As high energy implies huge gamma factors and fields close to $E_c$ this is also the genuine quantum regime. 

In this paper we will concentrate on the segment of the QED parameter space that has become accessible only recently, characterised by large intensities, $a_0 \gg 1$, and comparatively low energies, $\omega \ll mc^2$, typical for experiments with an all-optical setup. We will thus stay far below the Breit-Wheeler pair creation threshold and will have to consider a process that is not suppressed by either unfavourable powers or exponentials. A natural process that comes to mind is a crossing image of the Breit-Wheeler one, namely strong-field Compton scattering where a high-intensity beam of laser photons $\gamma_L$ collides with an electron beam emitting a photon $\gamma$. In this case one has to sum over all $n$-photon processes of the type
\be \label{NLC}
  e^- + n \gamma_L \to e^- + \gamma \; .
\ee
The study of this process(es) has a history almost as long as that of the laser. Intensity effects were addressed as early as 1963/64 in at least three independent contributions, by Nikishov, Ritus and Narozhnyi \cite{Nikishov:1963,Nikishov:1964a,narozhnyi:1964,Nikishov:1964b}, Brown and Kibble \cite{Brown:1964zz} and Goldman \cite{Goldman:1964}. These works are written from a particle physics perspective i.e.\ essentially by working out the relevant Feynman diagrams. For modern reviews of these development the reader is referred to \cite{McDonald:1986zz,Burke:1997ew}. Nikishov and Ritus in \cite{Nikishov:1964b} pointed out that $a_0^2$ is proportional to $E^2$ and hence the photon density $n_\gamma$. The precise relationship is
\be
  a_0^2 = \frac{\hbar e^2}{m^2c^2 \omega} \, n_\gamma = 4 \pi \alpha \nu^2 \, \lambdabar^3 n_\gamma \; ,
\ee
where $\nu \equiv \hbar\omega/mc^2$ is the dimensionless laser frequency and $\lambdabar^3 n_\gamma$ is the number of photons in a laser wavelength cubed. As the probability for the process (\ref{NLC}) is proportional to $a_0^{2n} \sim n_\gamma^n$ it becomes \textit{nonlinear} in photon density for $n>1$ and hence is called \textit{nonlinear Compton scattering} \cite{Nikishov:1964b}. Somewhat in parallel, the same process has been considered by the laser and plasma physics communities, with an emphasis, however, on the very low-energy and hence classical aspects. The appropriate notion is therefore nonlinear \textit{Thomson} scattering. These discussions were based on an analysis of the classical Lorentz-Maxwell equation of motion, typically using a noncovariant formulation and neglecting radiation damping. Some early references are papers by Sengupta \cite{Sengupta:1949}, Vachaspati \cite{Vachaspati:1962} and Sarachik and Schappert \cite{Sarachik:1970ap}. Since then there has been an enormously large number of papers from this perspective, many of which are quoted in the concise review \cite{Lau:2003}.

The main intensity effect can indeed be understood classically, the reason being the huge photon numbers involved, $\lambdabar^3 n_\gamma \simeq 10^{18}$ in a laser wavelength cubed. Due to the quiver motion in a (circularly polarised) plane wave laser field the electron acquires a quasi 4-momentum given by
\be \label{QUASIMOM}
  q \equiv p + \frac{ a_0^2 \, m^2}{2 k \cdot p} \, k \equiv p + q_L \; .
\ee
Hence, the electron acquires an additional, \textit{intensity dependent} longitudinal momentum $q_L$ caused by the presence of the laser fields. It may be obtained as the proper time average of the solution $p_\mu (\tau)$ of the classical equation of motion with $p_\mu = p_\mu(0)$ being the initial electron 4-momentum and $k_\mu = \omega n_\mu$ the lightlike 4-vector of the wave \cite{Kibble:1965zz}. Historically, (\ref{QUASIMOM}) was first found in the context of Volkov's solution \cite{volkov:1935} of the Dirac equation in a plane electromagnetic wave. Volkov explicitly wrote down the zero component $q^0$ while the generalisation (\ref{QUASIMOM}) seems to be due to Sengupta (note added at the end of his paper \cite{Sengupta:1952}; cf.\ also the textbook discussion in \cite[Chapter 40]{Berestetskii:1982}). Upon squaring $q$ one infers as an immediate consequence the intensity dependent mass shift,
\be \label{MASS-SHIFT}
  m^2 \to m_*^2 = m^2(1 + a_0^2) \; .
\ee
Although first predicted by Sengupta in 1952 \cite{Sengupta:1952} (see also \cite{Brown:1964zz,Kibble:1965zz}) it has so far never been observed directly \cite{McDonald:1999et}. A central topic of this paper will be to (re)assess the prospects for measuring effects due to the mass shift (\ref{MASS-SHIFT}).

The paper is organised as follows. We begin in Sect.~\ref{QED-sect} by reviewing the coherent state model of laser fields, which provides the link between classical laser light and light \textit{quanta} (photons) in quantum theory.  We then describe scattering amplitudes between these coherent states in QED, and how they are generated by an effective action describing interactions with a classical background field. We illustrate this theory with nonlinear Compton scattering, in  Sect.~\ref{NLC-sect}, and give a thorough discussion of the kinematics of the colliding particles. In Sect.~\ref{subsect:XS} we give a variety of predictions for both Lorentz invariant and lab--frame photon emission spectra. Our conclusions are presented in Sect.~\ref{Concs}.

\section{QED with classical background fields}\label{QED-sect}

We first address the question which asymptotic in--state we should take to describe the laser field. In principle, we would simply take the multi--particle state containing the appropriate number of photons of laser frequency and momentum, encoded in the 4-vector $k = (\omega, \vc{k})$. We are immediately faced with the problem of not knowing exactly how many photons are in the beam. Similarly, as we do not know how many photons will interact with, say, an electron during an experiment, we do not know what to take for the out--state. To overcome these problems we invoke the correspondence principle: due to the huge photon number in a high intensity beam it should be feasible to treat the laser  \emph{classically}, as some fixed background field. Formally, this is achieved by describing the laser beam, asymptotically, in terms of coherent states of radiation \cite{Schwinger:1953zza,Glauber:1962tt,Glauber:1963fi,Glauber:1963tx}. The coherent states have the usual exponential form
\be\label{codef}
	\ket{C} = \exp\sqrt{N}\int\!\frac{\ud^3k}{(2\pi)^3}\ C^\mu(\vc{k})\, a^\dagger_\mu(\vc{k})\,\ket{0}\;,
\ee
where $a^\dagger_\mu$ is the photon creation operator, $C_\mu(\vc{k})$ gives the (normalised) polarisation and momentum distribution of the photons in the beam and $N$ is the expectation value of the photon number operator (the average number of photons in the beam). As usual, the state is an eigenvector of the positive frequency part of $\hat A_\mu$, since
\be\label{codef2}
	a_\mu(\vc{k})\ket{C} = \sqrt{N}C_\mu(\vc{k})\ket{C},
\ee
Expanding the exponential in (\ref{codef}), we see that calculating S--matrix elements between states including coherent pieces is equivalent to a particular weighted sum over S--matrix elements of photon Fock states. Working with coherent states may also be thought of, physically, as neglecting depletion of the laser beam, i.e.\ taking the number of photons in the beam to remain constant \cite{BialynickiBirula:1973,Bergou:1980cp}. There is a natural connection between classical fields and coherent states as these states are the `most classical' available, having minimal uncertainty. The associated classical field is essentially, as we shall see, the Fourier transform of the distribution function $C$. To see this we turn to the calculation of S--matrix elements between coherent states.

Consider some scattering process with an asymptotic in--state containing the coherent state $C$, and some collection of other particles. For reasons which will shortly become clear, we will summarise all those particles \textit{not} in the coherent state by `in', so that our state is $\ket{\text{in};C}$. Similarly, we take an out state of the form $\bra{\text{out};C}$ where we have, in accord with the assumption of no beam depletion, the \emph{same} coherent state. In operator language, we are interested in calculating matrix elements $\bra{\text{out};C}\, \hat{\mathbb{S}}\, \ket{\text{in};C}$ of the S--matrix operator, 
\be \label{SOP}
	 \hat{\mathbb{S}} \equiv \mathcal{T} \exp\bigg[ -\frac{i}{\hbar} \int\limits_{-\infty}^\infty\!\ud t\  \hat{H}_I(t) \bigg]\;.
\ee
Here $\hat{H}_I(t)$ is the interaction Hamiltonian (in the interaction picture) and $\mathcal{T}$ denotes time--ordering. We now write the coherent state (\ref{codef}) as a translation of the vacuum state (see e.g.\ \cite{Gottfried:2004}),
\be
	\ket{C} = \hat{T}_C \, \ket{0}\;,
\ee
where the commutator of the translation operator and the photon annihilation operator is
\be
	\big[ \hat{a}_\mu(\vc{k})\, , \, \hat{T}_C \big] = C_\mu(\vc{k})\, \hat{T}_C \;.
\ee
Extracting the translation operator from the states\footnote{Under the usual assumption of no forward scattering. For the photons, this requires $C_\mu(\vc{k}')=0$ for any scattered photons of momentum $\vc{k}'$.}, we are left with ordinary asymptotic Fock states but with a modified S--matrix operator,
\be\label{S-rule0}
	\bra{\text{out};C}\, \hat{\mathbb{S}}\, \ket{\text{in};C} = \bra{\text{out}}\,\hat{T}^{-1}_C \, \hat{\mathbb{S}} \, \hat{T}_C \, \ket{\text{in}}\;.
\ee
From the definition (\ref{SOP}) of $\hat{\mathbb{S}}$, the effect of the translation operators is to shift any photon operator $\hat A_\mu$ appearing in the interaction Hamiltonian by (the Fourier transform of) $C_\mu(\vc{k})$ which we denote by $\mathcal{A}_\mu (x)$. Hence, the fermions interact with the full quantum photon field $\hat A_\mu$ and a classical background field, $\mathcal{A}_\mu (x)$.

To be precise, and switching to the more common LSZ language, S--matrix elements are given by the on-shell Fourier transform of Feynman diagrams with amputated external legs, as usual, but where the Feynman diagrams are generated by the action
\be\label{action}
	S[A,\mathcal{A},\psi,\overline\psi] = \int\!\ud^4x\ -\frac{1}{4}F_{\mu\nu}F^{\mu\nu} +  \overline\psi(i\big[\slashed{\partial}+ie\slashed{\mathcal{A}}+ie\slashed{A}\big]-m)\psi\;.
\ee
This is almost the ordinary QED action, but the photon field in the interaction term is shifted by $\mathcal{A}_\mu$, explicitly given by
\be\label{co-cl}
	\mathcal{A}_\mu(x) = \sqrt{N} \int\!\ud^3k\ \frac{\e^{-ik\cdot x} }{(2\pi)^{3/2}\sqrt{2|\vc{k}|}} \ C_\mu(\vc{k}) + \text{c.c.}\; \bigg|_{k^2=0}\;.
\ee
This potential gives the classical electromagnetic fields associated with the momentum distributions $C_\mu(\vc{k})$. Note that only the interaction terms of the action are affected by the presence of the background field, following (\ref{S-rule0}). We therefore have a natural and quite elegant way to calculate -- we do not need to directly add up the individual contributions of the infinite series of terms generated by expanding the asymptotic coherent state. Instead, we simply include a classical background in the action which contains all the information about the chosen asymptotic photon distributions. Following \cite{Kibble:1965zz,Frantz:1965} these results can be summarised by
\be\label{S-rule}
  \bra{\text{out};C}\,\hat{\mathbb{S}}\,\ket{\text{in};C} =  \bra{\text{out}}\,\hat{T}^{-1}_C \, \hat{\mathbb{S}} \, \hat{T}_C \, \ket{\text{in}} \equiv   \bra{\text{out}}\,\hat{\mathbb{S}}[\mathcal{A}]\,\ket{\text{in}}\;,
\ee
where, on the right hand side, the asymptotic states are ordinary particle number states, with no coherent pieces, and the photon fields in the S--matrix operator are translated by $\mathcal{A}_\mu$.

Briefly, the same result can be recovered entirely in the path integral, or functional, language, following e.g. \cite[Chapter~9.2]{Weinberg:1995mt}. The construction of S--matrix elements between coherent states proceeds just as for elements between Fock states, but the asymptotic vacuum wavefunctional must be replaced by coherent state wavefunctionals. Ordinarily it is the vacuum which is responsible for introducing the $i\epsilon$ prescription into the action and from there into the field propagators. A coherent state does this and more -- it translates the photon field in the interaction terms by the classical field (\ref{co-cl}), recovering (\ref{S-rule}).

\begin{figure}[!!ht]
\centering\includegraphics[width=0.9\textwidth]{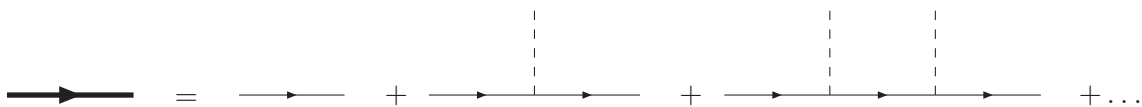}
\caption{\label{Evolk} Perturbative expansion of the electron propagator in a background field.}
\end{figure}

Note that the modified action (\ref{action}) remains quadratic in the fermion field. All effects of the background are therefore contained in a modification of the electron propagator. The result is that, in Feynman diagrams, the propagator becomes `dressed' by the background field $\mathcal{A}_\mu$ which surrounds the electrons. The propagator will be represented by a heavy line as in Fig.~\ref{Evolk}, and has a perturbative expansion in terms of a free electron propagator interacting an infinite number of times with $\mathcal{A}_\mu$, as represented by the dashed line.
The Feynman rules of the theory are otherwise unchanged from QED -- there is a single three field vertex which joins the photon propagator and two of the dressed fermion propagators. This background field approach is equivalent to adopting a Furry picture \cite{Furry:1951zz}, in which the `interaction' Hamiltonian describes the quantum interactions while the interaction with the background $\mathcal{A}_\mu$ is treated as part of the `free' Hamiltonian.

In general, the fermion propagator will have no closed form expression. Since an intense background will be characterised by numbers larger than one (such as the intensity parameter $a_0$), a perturbative expansion in the background is not suitable. We can of course use a coupling expansion, but this leaves us with an infinite number of Feynman diagrams to calculate for \emph{any} process, even at tree level. Fortunately, for the backgrounds considered in this paper and discussed below, the electron propagator is known exactly, allowing us to treat the background field exactly. We will now illustrate these ideas by applying them to the process of interest in this paper; nonlinear Compton scattering.

\section{Nonlinear Compton scattering}\label{NLC-sect}

\begin{figure}[!ht]
\centering\includegraphics[scale=0.7]{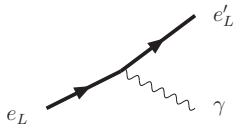}
\caption{Nonlinear Compton
scattering Feynman diagram using dressed electrons (subscript $L$).}
\label{NLC-v-pic}
\end{figure}

In this process an electron, incident upon a laser, scatters a photon out of the beam. Using the background field approach described above, we use the action (\ref{action}), which contains the effects of the laser, and take the asymptotic in-- and out--states to be, respectively,
\be\label{states}
	\ket{\vc{p},\lambda}\;,\qquad \bra{\vc{p}',\lambda';\vc{k}',\epsilon}\;.
\ee
The pair $(\vc{p},\lambda)$ give the momentum and spin state of the incoming electron, similarly $(\vc{p}',\lambda')$ describe the outgoing electron and $(\vc{k}',\epsilon)$ are the momentum and polarisation tensor of the scattered photon.  Only one Feynman diagram contributes to this process at tree level, shown in Fig.~\ref{NLC-v-pic}. Note that the analogous scattering amplitude with `naked' electrons, corresponding to spontaneous photon emission in vacuum, vanishes due to momentum conservation.
Calculating the corresponding S--matrix element amounts to amputating the external legs and integrating over the single vertex position. Amputating and Fourier transforming the electron propagator in a background field gives us the solutions of the Dirac equation in that background \cite{Nikishov:1963,Brown:1964zz,Goldman:1964,Kibble:1965zz}. We will write these electron wavefunctions as $\Psi_{\ssvc{p}\,\lambda}(x)$. The S--matrix element of the process in Fig.~\ref{NLC-v-pic} therefore reduces to
\be\label{tocalc}
	\bra{\vc{p}',\lambda';\vc{k}',\epsilon}\,\hat{\mathbb{S}}[\mathcal{A}]\,\ket{\vc{p},\lambda}=-ie \int\!\ud^4x\ \overline\Psi_{\ssvc{p}'\,\lambda'}(x)\ \frac{\e^{ik'\cdot \,x}}{\sqrt{2|\vc{k}'|}}\slashed\epsilon\ \Psi_{\ssvc{p}\,\lambda}(x)\;.
\ee
To proceed we need to pick a background field so that we can explicitly calculate the wavefunctions $\Psi_{\ssvc{p}\,\lambda}(x)$ and therefore the S--matrix element (\ref{tocalc}). This is the focus of the next section.

\subsection{Plane waves and Volkov electrons}

We will model the laser by a plane wave, $\mathcal{A}_\mu\equiv \mathcal{A}_\mu(k\cdot x)$, with $k$ a lightlike four--vector characterising the laser beam direction. The electron wavefunctions in such a background, or `Volkov electrons' \cite{volkov:1935}, are known exactly. The propagator is also known and may be derived either in field theory or using a first quantised (proper time) method \cite{Schwinger:1951nm}. For a textbook discussion see \cite[Chapter 40]{Berestetskii:1982}. The Volkov electron is
\be\label{wave}
	\Psi_{\ssvc{p}\lambda}(x) = e^{-ip\cdot x}\exp\bigg\{\frac{1}{2ik\cdot p} \int\limits^{k\cdot x}\!\ud\xi\ 2e p\cdot\mathcal{A}(\xi)-e^2 \mathcal{A}^2(\xi)\bigg\}\bigg[1+\frac{e}{2k\cdot p}\slashed{k}\slashed{\mathcal{A}}\bigg]u_{\ssvc{p}}\;,
\ee
where $p^2=m^2$ and $u_{\ssvc{p}}$ is the usual electron spinor.

To better understand this wavefunction we specialise from here on to the case of $\mathcal{A}_\mu$ being a circularly polarised plane wave of amplitude $a$,
\be
	\mathcal{A}^\mu = a_1^\mu \cos(k\cdot x) + a_2^\mu\sin(k\cdot x)\;,
\ee
where $a_j\cdot k=0$ and $a_j\cdot a_k=-a^2\delta_{jk}$. The electron wavefunction becomes
\be\label{wave1}
	\Psi_{\ssvc{p}\lambda}(x) = \exp\bigg[-iq\cdot x-ie\frac{a_1\cdot p}{k\cdot p}\,\sin(k\cdot x)+ie\frac{a_2\cdot p}{k\cdot p}\,\cos(k\cdot x)\bigg]\times \ldots
\ee
We have not given the explicit form of the spinor part; it is easily written down and not needed for the discussion in this section. The important effect is that the electron acquires the quasi 4--momentum $q$ defined in (\ref{QUASIMOM}) from the laser field with the intensity parameter $a_0$  given by
\be
   a_0^2 \equiv \frac{e^2 a^2}{m^2}\;.
\ee
Technically, the origin of the quasi--momentum lies in a separation of the exponent in (\ref{wave}) into a Fourier zero mode and oscillatory pieces, with the zero mode causing the momentum shift, $p \to q$. Inserting the wavefunctions (\ref{wave1}) into (\ref{tocalc}), and omitting the details of the calculation \cite{narozhnyi:1964}, we find that the scattering amplitude is a periodic function with Fourier series
\be\label{sum}
	\bra{\vc{p}',\lambda';\vc{k}',\epsilon}\,\hat{\mathbb{S}}[\mathcal{A}]\,\ket{\vc{p},\lambda}= \frac{1}{(2|\vc{k}'|\  2E_{\ssvc{q}'}\ 2E_{\ssvc{q}})^{1/2}}\sum\limits_{n=1}^\infty M(n)\ \delta^{(4)}\left(q+n k-q'-k'\right)\;.
\ee
A discussion of the amplitudes $M(n)$ may be found in \cite[Chapter 100]{Berestetskii:1982}. We will give below the explicit form of the squared amplitudes summed over spins $\lambda$, $\lambda'$ and polarisations $\epsilon$. We do not consider polarised scattering and angular distributions in this paper, though these topics are interesting in themselves and are discussed in, for instance, \cite{Tsai:1992ek,Esarey:1993zz,Ivanov:2004fi}.

\begin{figure}[ht!!]
\begin{center}
\includegraphics[scale=0.6]{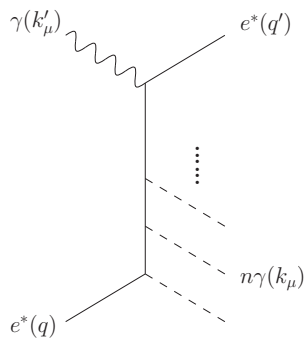}
\caption{\label{fig:NLC} The effective Feynman diagram describing the $n^\text{th}$ harmonic process; an electron of mass $m_*$ absorbs $n$ laser photons of momentum $k_\mu$ and emits a photon of momentum $k'_\mu$.}
\label{Fig:NLC}
\end{center}
\end{figure}

The sum in (\ref{sum}) is not a coupling expansion, nor does it appear directly from an expansion of the coherent state into Fock states. Instead, the momentum--conserving delta function in the $n^\text{th}$ term implies that $M(n)$ can be identified with the amplitude for an electron of momentum $q$ and mass $m_*$, absorbing $n$ photons of momentum $k$ and emitting one scattered photon of momentum $k'$,
\be\label{effec}
	e_*(q) + n\gamma(k) \to e_*(q')+ \gamma(k')\;,
\ee
as illustrated in Fig.~{\ref{Fig:NLC}}.
As pointed out in the introduction, these multi--photon processes are the origin of the name `nonlinear' Compton scattering. It is simplest to use the language of quasi momenta to formulate the  kinematics of (\ref{sum}) as (\ref{effec}) is a process involving effective particles. The asymptotic particle kinematics may be reconstructed from the relation (\ref{QUASIMOM}) between $p$ and $q$. The processes with $n>1$ correspond to higher harmonics. Note that the $n=1$ process is analogous to ordinary, `linear' Compton scattering. It is possible to normalise such that one does indeed recover the Compton cross section at $a_0 = 0$. We will use this below as a reference cross section for experimental signals.

\subsection{Kinematics -- forward and back scattering}\label{subsect:KINEMATICS}

We will now study the kinematics implied by the momentum conservation in (\ref{sum}), finding an expression for the emitted photon frequency in terms of incoming particle data which generalises the standard Compton formula for the photon frequency shift. This will later be used when we predict the emitted photon spectrum.

The delta function in (\ref{sum}) implies the momentum conservation equation
\be\label{new-cons}
  q + n k =q' + k' \; ,
\ee
where $q$ is given by (\ref{QUASIMOM}), $q'$ being defined analogously with $p$ replaced by $p'$. As $k$ is light-like we have
\be \label{QDOTK}
  q \cdot k = p \cdot k \; , \quad q' \cdot k = p' \cdot k \; .
\ee
It is useful to first discuss the kinematics in terms of the Mandelstam invariants
\bea
  s_n &=& (q + n k)^2 = m_*^2 + 2n k \cdot p \ge m_*^2 \; , \label{S} \\
  t_n &=& (n k - k')^2 = - 2n k \cdot k' \le 0 \; , \label{T} \\
  u_n &=& (n k - q') = m_*^2 - 2n k \cdot p' \; . \label{U}
\eea

\begin{figure}[!ht]
\begin{center}
\hspace{2cm}
\includegraphics[scale=1.2]{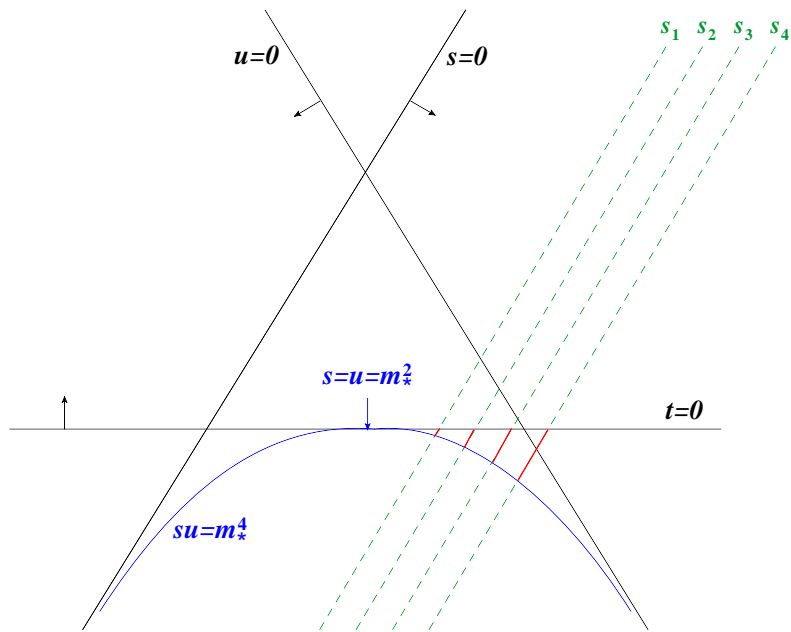}
\caption{\label{Fig:mandelstam}(color online) Mandelstam plot for nonlinear Compton scattering. Solid segments of dashed lines correspond to allowed $u_n$ and $t_n$ regions for each depicted value of $s_n$.}
\end{center}
\end{figure}

Recall that these are not independent as $s_n + t_n + u_n = 2 m_*^2$. As each of them depends on the photon number $n$ they will be different for each of the sub-processes (\ref{effec}). The physically allowed parameter ranges are displayed in the Mandelstam plot of Fig.~\ref{Fig:mandelstam}. For the $n$-photon sub-process, if $s = s_n$ is held fixed, the allowed $t$ and $u$ ranges (highlighted in red/full segments of dashed lines) are
\be \label{TRANGE}
  \begin{array}{lll}
  t_{\mathrm{min}} = 2 m_*^2 - s_n - m_*^4/s_n \qquad & u_{\mathrm{max}} = m_*^4/s_n \qquad & \mbox{back scattering} \\
  t_{\mathrm{max}} = 0 \qquad & u_{\mathrm{min}} = 2m_*^2 - s_n \qquad & \mbox{forward scattering}
  \end{array}
\ee
Obviously, the allowed $t$-range increases with photon number $n$.

In order to find the generalisation of Compton's formula for the scattered photon frequency (thus abandoning manifest covariance) we square (\ref{new-cons}) so that we may remove $q'$ from the game via
\be
  n\, k \cdot q = k' \cdot q' = q \cdot k' + n\, k \cdot k'\;,
\ee
where the second equality follows directly from (\ref{new-cons}). Using the definition (\ref{QUASIMOM}) and (\ref{QDOTK}) we trade $q$ for $p$, arriving at an equation in terms of the asymptotic, on--shell momenta
\be\label{hi-a0}
  n\, k \cdot p = k'\cdot p + \bigg(n+a_0^2 \, \frac{m^2}{2k \cdot p}\bigg)\, k \cdot k' \;,
\ee
where $k'^2=0$ and $p^2=m^2$. We will assume, in what follows, that the electron and laser meet in a head on collision. That is, incident momenta are
\be
	k^\mu = \omega(1, \vc{n}) \,, \qquad p^\mu = (\Ep, -|\vc{p}| \, \vc{n}) \; , \qquad |\vc{n}|=1 \; .
\ee
Primed (outgoing) quantities are defined analogously. For a head on collision the only angle in play is the standard scattering angle $\theta$ of the photon, determined via
$\vc{n} \cdot \vc{n}' \equiv \cos \theta$. The remaining scalar products become
\be
  \vc{n} \cdot \vc{p} = -|\vc{p}| \; , \quad \vc{n}' \cdot
  \vc{p} = -|\vc{p}| \cos \theta \; .
\ee
From now on we measure all energies in units of the (bare) electron mass, $m$. This introduces the dimensionless parameters
\be \label{GAMMABETA}
  \nu \equiv \frac{\omega}{m} \;, \quad \gamma \equiv \frac{\Ep}{m} \equiv \cosh \zeta \;, \quad \beta \gamma \equiv \frac{|\vc{p}|}{m} \equiv \sinh \zeta \; ,
\ee
where $\zeta$ is the rapidity such that
\be
  \beta \equiv \frac{|\vc{p}|}{\Ep} = \sqrt{1 - 1/\gamma^2} \equiv \tanh \zeta \; .
  \label{BETA}
\ee
Of course, $\beta$ and $\gamma$ are the usual Lorentz factors characterising the frame of reference from the electron's point of view. $\beta = 0$, for instance, corresponds to the (asymptotic) electron rest frame. Using these definitions, equation (\ref{hi-a0}) may be rearranged to express the intensity dependent scattered photon frequency as
\be \label{NUPRIME}
  \nu'_n (\theta) = \frac{n \nu}{1 + \kappa_n(a_0) \,  e^{-\zeta} \, (1-\cos\theta)} \; .
\ee
Here, $e^{-\zeta}$ is the (inverse) Doppler shift factor for a head-on collision,
\be \label{DOPPLER}
  e^{-\zeta} = \gamma(1-\beta) = \sqrt{\frac{1-\beta}{1 + \beta}}\; .
\ee
Going back to (\ref{NUPRIME}) we see that all the intensity dependence resides in the coefficient
\be \label{KAPPA}
  \kappa_n (a_0) \equiv  n\nu - \beta \gamma + a_0^2 \,\gamma (1-\beta)/2 = n\nu - \sinh \zeta + a_0^2 \, e^{-\zeta}/2  \; .
\ee
Standard (`linear') Compton scattering is reobtained by setting $n = 1$ and $a_0 = 0$ (no intensity effects). In this case (\ref{NUPRIME}) and (\ref{KAPPA}) give back the ordinary Compton formula,
\be \label{LINCOMP}
  \nu_1'   =  \frac{\nu}{1 + (\nu - \beta \gamma) \, \gamma (1 -\beta) \, (1-\cos\theta)} = \frac{\nu}{1 + (\nu - \sinh \zeta) \, e^{-\zeta} \, (1 - \cos \theta)} \; .
\ee
So, technically speaking, the two intensity effects on the scattered frequency are the replacements (i) $\nu \to n\nu$ in the numerator and (ii) $\kappa_1(0) \to \kappa_n (a_0)$ in the denominator. Explicitly, the latter is
\be \label{REPLACE}
  \nu - \beta \gamma \longrightarrow n\nu - \beta \gamma + a_0^2 \,\gamma (1-\beta)/2  \;,  \qquad \mbox {or} \qquad \nu - \sinh \zeta \longrightarrow n\nu - \sinh \zeta + a_0^2 \, e^{-\zeta}/2 \; .
\ee
The possibility of the incoming electron absorbing $n>1$ laser photons may be interpreted, in a classical picture, as the generation of the $n^\text{th}$ harmonic, modulated by both relativistic and intensity effects. Using a \textit{linearly} polarised beam the first few harmonics have indeed been observed experimentally by analysing the photon distribution as a function of azimuthal angle, $\phi$. The second and third harmonics have clearly been identified from their quadrupole and sextupole radiation patterns \cite{Chen:1998}.

For each harmonic number $n$, the allowed range of scattered photon frequencies $\nu'_n$ is finite. The boundary values of this interval (which is the $t$-interval in the Mandelstam plot Fig.~\ref{Fig:mandelstam}) correspond to forward and back scattering at $\theta = 0$ and $\pi$ respectively,
\be \label{BS1}
  \nu'_n(0) = n\,\nu\;,\qquad \nu'_n (\pi) = \frac{n \nu}{1 + 2 \kappa_n(a_0) \,  e^{-\zeta}}\;.
\ee
The assignment of minimum and maximum depends on the sign of $\kappa_n$,
\be \label{INTERVALS}
  \begin{array}{ccccccc}
  \kappa_n > 0 &\implies\quad& \nu'_n (\pi) < \nu'_n (\theta) < n\,\nu &\quad& \mbox{red shift}  &\quad& \mbox{`Compton'} \\[5pt]
  \kappa_n < 0 &\implies\quad& n\,\nu < \nu'_n (\theta) < \nu'_n (\pi) &\quad& \mbox{blue shift} &\quad& \mbox{`Inverse Compton'}
  \end{array}
\ee
So, if $\kappa_n > 0$, the allowed scattered photon energies $\nu'_n$ are red shifted relative to $n\nu$, the energy of the $n$ absorbed laser photons. This clearly includes the case $a_0=0$, $\gamma=1$ and $n=1$ which describes Compton's original scattering experiment in the electron rest frame. In accelerator language, this case sees the laser fired onto a fixed electron target; the laser photon transfers energy to the target, so that the scattered photon is red-shifted ($\nu' < n\nu$).

On the other hand, if $\kappa_n < 0$, the scattered photon's energy is blue shifted from $n\nu$. The situation when the photon gains energy from the electrons is often referred to as `inverse' Compton scattering. This is of relevance in astrophysics, for instance in the Sunyaev-Zeldovich effect \cite{Sunyaev:1970er, Sunyaev:1980vz, Birkinshaw:1998qp}.  A particularly simple and important scenario is provided by the backscattering of the laser pulses, $\theta = \pi$, in the high energy limit (inverse Compton regime).  We take $\gamma \gg 1$ so that $e^\zeta \simeq 2\gamma$ and we assume $\kappa_n< 0$, whereupon the scattered frequency becomes, from (\ref{NUPRIME}),
\be \label{BS2}
  \nu'_n (\pi) = \frac{n\nu \, e^{2\zeta}}{1 + a_0^2 + 2n\nu \, e^{\zeta}} \simeq \frac{4 \gamma^2 n\nu}{1 + a_0^2 + 4 \gamma n \nu} \; ,
\ee
where the approximation is valid for high energy. In this regime one may distinguish between two different limits,
\bea
  \nu'_n (\pi) &=& 4 \gamma^2 n\nu/a_0^2 \quad \mbox{if} \quad  4 \gamma n \nu \ll 1 \ll a_0^2 \label{NUPRIME-HE} \\
  \nu'_n (\pi) &=& \gamma \hspace{1.25cm} \quad \mbox{if} \quad 1 + a_0^2 \ll 4 \gamma n \nu
\eea
It is the former subcase which is typically realised\footnote{SLAC E-144 had $\gamma\nu = O(1)$ so all terms in the denominator of (\ref{BS2}) were of comparable magnitude.} for optical photons ($\nu \simeq 10^{-6}$) and moderate values of harmonic number $n$. Thus, as long as $a_0 \lesssim 2 \gamma$, the back scattered frequency (\ref{NUPRIME-HE}) is (i) blue-shifted with respect to the incoming $n^\text{th}$ harmonic frequency $n\nu$ and (ii) for $n=1$, red-shifted compared to the linear `kinematic edge' (the maximal, back-scattered frequency, $\nu_{\mathrm{max}}^\prime$) as emphasised already by McDonald \cite{McDonald:1986zz}. Explicitly, this red-shift is
\be
    4 \gamma^2 \nu \longrightarrow 4 \gamma^2 \nu/a_0^2 \; , \quad \gamma \gg 1 \; , \quad 4 \gamma n \nu \ll 1 \ll a_0^2 \; .
\ee
From the definition of $\kappa_n$ given in (\ref{KAPPA}) it is clear that, given any fixed experimental setup  (i.e.\ incoming electron energy and intensity parameters $\zeta$ and $a_0$), $\kappa_n$ will eventually become positive, and remain so for all higher harmonics with
\be\label{FLOOR}
	n > \left\lfloor\frac{\sinh\zeta - a_0^2e^{-\zeta}/2}{\nu}\right\rfloor \equiv n_0\; ,
\ee
where $\lfloor b \rfloor$ denotes the nearest integer less than or equal to $b$. Thus, for a given experimental setup, scattered photons corresponding to harmonic generation with $n > n_0$ can only have energies red--shifted relative to the energy $n\nu$ absorbed by the electron.  Alternatively, we can fix $n$ and so define a critical intensity, from the vanishing of $\kappa_n$, which allows us to tailor the emission spectrum. The critical intensity parameter is
\be\label{a0-crit}
	a_{0,\mathrm{crit}}^2 (n) \equiv 2 \gamma (1+ \beta)(\beta \gamma - n\nu) = 2 e^\zeta (\sinh \zeta - n\nu) = e^{2\zeta} - 2n\nu \, e^\zeta - 1  \stackrel{!}{\ge} 0 \; .
\ee
For $a_0 = a_{0,\mathrm{crit}}(n_0)$ all harmonics with $n>n_0$ ($n < n_0$) will be red-shifted (blue-shifted). For the extreme choice of $n_0 = 1$, \textit{all} scattered frequencies will be red-shifted for intensities above $a_{0,\mathrm{crit}}(1)$, as in, for example, fixed target mode ($\gamma = 1$). We are, however, more interested in the colliding mode (high energy). Then, for $\gamma \gg 1$, we can approximate $a_{0,\mathrm{crit}}^2$ from (\ref{a0-crit}) as
\be \label{a0-crit-approx_n}
  a_{0,\mathrm{crit}}^2 \simeq 4 \gamma^2 - 4\gamma n\nu  \; .
\ee
When $4 \gamma n \nu \ll 1 $ as above, $a_{0,\mathrm{crit}}$ becomes effectively $n$-independent
\be \label{a0-crit-approx}
  a_{0,\mathrm{crit}} \simeq e^\zeta \simeq 2\gamma \; .
\ee
As a numerical example consider the facility at the Forschungszentrum Dresden-Rossendorf (FZD) with a 100 TW laser and a 40 MeV linac \cite{FZD:2009}.  This implies $\gamma = 80$, $\nu = 2\times 10^{-6}$ and $a_0 \simeq 20$, so that all harmonics are relatively blue shifted up to $n\simeq 3.9\times 10^7$ -- as we will see, emission rates at this $n$ are basically zero. In this case, the critical value of $a_0$, above which all harmonics ($n \ge 1$) are relatively red shifted compared to $n\nu$, is $a_0 \gtrsim 2 \gamma = 160$, an order of magnitude above the expected available intensity. One may verify, for example, that for $a_0 = 200$, $\kappa_n > 0$ for all $n$.

The discussion above will be illustrated in the next section when we discuss the photon spectra as a function of scattered frequency, $\nu'$. In particular, we will see that, even if  backscattering does not necessarily maximise the scattered photon frequency, it nevertheless gives us the strongest signal for which to search experimentally, namely the red-shift of the Compton edge (parameters permitting).

To better understand the different behaviours of the harmonics, it is useful to write $\kappa_n$ in terms of lab frame variables. For a head-on collision (which we assume), say along the $z$-axis, all momenta involved are longitudinal. The total 3-momentum, call it $\vc{P}$, is then given by
\be \label{TOTALMOM}
   \vc{P} \equiv n\vc{k} + \vc{q} = n \vc{k} + \vc{p} + \vc{q}_L = m \left( n\nu - \sinh \zeta + a_0^2 \, e^{-\zeta}/2   \right) \hat{\vc{z}}
    = m\, \kappa_n \, \hat{\vc{z}} \; .
\ee
The lab-frame physics involved in a head-on collision ($\vc{p} = -(\beta\gamma/\nu)\vc{k}$) depends crucially on the relative magnitude of the three terms contained in $\kappa_n$,
\bea
  n |\vc{k}|/m &=&  n\nu \; , \\
  |\vc{p}|/m &=&  \sinh \zeta \; , \\
  |\vc{q}_L|/m &=& a_0^2\, e^{-\zeta}/2 \; .
\eea
Consider again Compton's original experiment with an electron at rest and $a_0 = 0$.  This corresponds to $\vc{q}_L = \vc{p} = 0$, so the only 3--momentum is that of the single incoming photon which delivers part of its energy to the electron and hence is red-shifted. If we now increase the electron energy in the lab (using a standard or wake field acceleration scheme) this red-shift turns into a blue-shift ($\nu' > \nu$) as soon as $|\vc{p}| > |\vc{k}| = m\nu$. This happens exactly where the total momentum, $\vc{P} = \vc{k} + \vc{p}$, changes direction from pointing in direction $\vc{k}$ to $-\vc{k}$. Hence, at this particular point $\vc{P}$ passes through zero, which, of course, defines the centre-of-mass (CM) frame where there is no frequency shift at all, $\nu' = \nu$.

If we now turn on intensity ($a_0 > 0$) the total momentum acquires an additional, laser induced, contribution $\vc{q}_L$ along $\vc{k}$. So, in fixed target mode large intensity will result in a significant enhancement of the Compton red-shift.  If, on the other hand, we assume colliding mode with a blue-shift at $a_0 = 0$, then the $\vc{q}_L$ contribution in $\vc{P}$ works against the `influence' of $\vc{p}$. As a result, the blue shift $\nu' > \nu$ at zero intensity is reduced, resulting in a red-shift of the kinematical Compton edge ($\nu'_{\mathrm{max}}$). If $a_0$ is large enough this latter red-shift may completely cancel the inverse Compton blue-shift. Again, this happens when the total momentum $\vc{P} = \vc{k} + \vc{p} + \vc{q}_L$ vanishes ($\kappa_1 = 0$) i.e.\ in the `CM frame' which is now an intensity dependent notion as $\vc{q}_L$ depends on $a_0$.

If we finally allow for higher harmonics $n>1$, with the total momentum becoming $\vc{P} = n\vc{k} + \vc{p} + \vc{q}_L$, we can balance $\vc{p}$ by increasing $a_0$ or $n$ or both. The transition point, $\kappa_n = 0$, defines  a `CM frame' for the $n^\text{th}$ process.  At this point, the range of the $n^\text{th}$ allowed harmonic collapses to a point, $\nu'_n(\theta) = n\nu$, as the $\theta$ dependence in (\ref{NUPRIME}) drops out. Strictly speaking, this can only occur for at most one value of $n$, but neighbouring $n$'s will still have rather small spectral ranges (see Fig.~\ref{Fig:MOVIE} below).

\section{Photon emission rates}\label{subsect:XS}

\subsection{Lorentz invariant characterisation}

The $S$-matrix element represented by the Feynman diagram of Fig.~\ref{NLC-v-pic}, and given implicitly in (\ref{sum}) may readily be translated into an emission rate \cite{Nikishov:1963,Berestetskii:1982}. The non--trivial contribution to the differential rate for emitting a photon of frequency $\omega' = m\nu'$ per unit volume per unit time, in the $n^\text{th}$ harmonic process, i.e. the process (\ref{effec}), comes from the differential probability\footnote{We normalise such that we recover the Klein-Nishina cross section for linear Compton scattering for $n=1$ as $a_0\to0$, see e.g.\ \cite{Berestetskii:1982}.} \cite{Berestetskii:1982}
\be
  \frac{\ud W_n}{\ud x} = \frac{1}{(1+x)^2} \, \mathfrak{J}_n (z(x)) \; ,
  \quad n \ge 1 \; ,
\ee
where $x$ is the dimensionless Lorentz invariant
\be \label{X}
  x \equiv \frac{k \cdot k'}{k \cdot p'} = \frac{t_n}{u_n - m_*^2} \ge 0 \; .
\ee
The kinematically allowed range for $n^\text{th}$ harmonic generation is given by the interval
\begin{align}
  \label{temp} 0 &\le x \le y_n \; ,\\
  \label{YN} y_n &\equiv \frac{2n \, k \cdot p}{m_*^2} = \frac{s_n}{m_*^2} - 1 \ge 0 \; ,
\end{align}
which corresponds to the $t$-range given in (\ref{TRANGE}), highlighted in Fig.~\ref{Fig:mandelstam}. The endpoints $x = y_n$ are located on the hyperbola $su = m_*^4$.  For $x$ outside of this range the $n^\text{th}$ partial rate vanishes.

\begin{figure}[!ht]
\begin{center}
\includegraphics[scale=0.7]{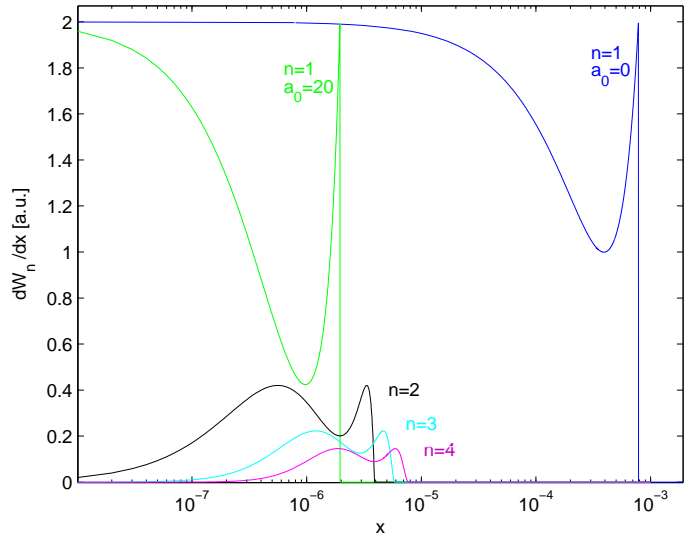}
\caption{\label{Fig:PartialRates20}(Color online) Partial emission rates ($n=1\ldots 4$) for nonlinear Compton scattering as a function of the Lorentz invariant $x$ at intensity $a_0 = 20$, compared to linear Compton scattering ($a_0=0$ curve). Horizontal log scale.}
\end{center}
\end{figure}

The function $\mathfrak{J}_n$ is
\be
  \mathfrak{J}_n (z) \equiv - \frac{4}{a_0^2} J_n^2 (z)
  + \left( 2 + \frac{x^2}{1+x} \right)
  \Big[  J_{n-1}^2 (z) + J_{n+1}^2 (z) - 2 J_n^2 (z) \Big] \; ,
\ee
the $J_n$ being Bessel functions of the first kind. Their argument is another Lorentz invariant
\be \label{Z}
  z(x) \equiv \frac{2a_0}{y_1} \sqrt{\frac{x(y_n - x)}{1 + a_0^2}} \; .
\ee
Both upper and lower limits of $x$ correspond to $z=0$ and hence zeros of $\mathfrak{J}_n (z)$ for all $n > 1$. The first few partial emission rates for $\Ep = 50$ MeV, $\omega = 1$ eV (hence $\gamma = 10^2$, $\nu = 2 \times 10^{-6}$) and $a_0=20$ are plotted in Fig.~\ref{Fig:PartialRates20}. Linear Compton ($a_0=0$ and $n=1$) data is presented for comparison.

The figure clearly shows the appearance of higher harmonics ($n>1$) with, however, a reduced signal strength as compared to the fundamental frequency. Writing the Compton edge (\ref{YN}) as
\be
  y_n = y_n (a_0) = y_1(0) \, \frac{n}{1+a_0^2} \; ,
\ee
where $y_1(0)$ corresponds to linear Compton scattering, we see that the edge $x = y_1 (a_0)$ of the first harmonic will always be shifted to the left by a factor $1/(1+a_0^2)$. The same is true for the higher harmonics until $n > 1 + a_0^2$. For $a_0 \gg 1$ these large harmonics will, however, be invisible due to their very small signal strength.

\begin{figure}[!ht]
\centering\includegraphics[scale=0.7]{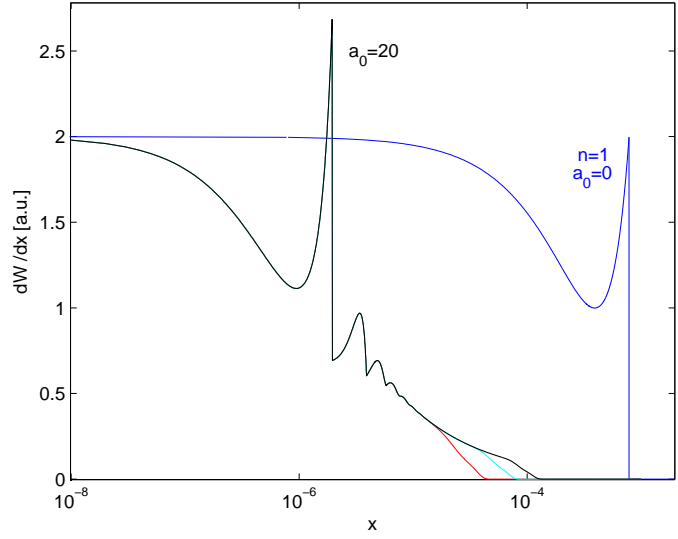}
\caption{\label{Fig:ratesum}(Color online) Sum of partial emission rates from $n = 1 \ldots 30$ (red, lower curve), $60$ (cyan, middle curve) and $100$ (black, top curve) for nonlinear Compton scattering (head-on collision) at intensity $a_0 = 20$. The curves are indistinguishable for $x$ $\simleq$ $10^{-5}$. Linear Compton data (blue, $n=1$, $a_0=0$) added for comparison.	}
\end{figure}

To obtain the total rate one just sums over photon numbers $n$, i.e.\ over all harmonics,
\be
  \frac{\ud W}{\ud x} = \sum_{n=1}^\infty \frac{\ud W_n}{\ud x} \; ,
\ee
where it is understood that the $n$-th term is supported on $0 \le x \le y_n$, with $x$ given in (\ref{X}). The partial sums up to $ n = 30$, $60$ and $100$ are shown in Fig.~\ref{Fig:ratesum}, along with the linear Compton spectrum. Again we note the significant shift of the fundamental Compton edge at $x = y_1 (a_0)$ together with side maxima due to the higher harmonics.   Interestingly, the fundamental ($n=1$) signal gets amplified due to superposition of the higher harmonic rates from Fig.~\ref{Fig:PartialRates20}. This suggests that, for $a_0 > 1$, the signal to noise ratio may become larger than for the linear case, while the full width at half maximum may become smaller. By tuning $a_0$ to an optimal value one may thus design X-rays of a given frequency and width.

\subsection{Lab kinematics: energy dependence}

\begin{figure}[!ht]
\begin{center}
\includegraphics[scale=0.7]{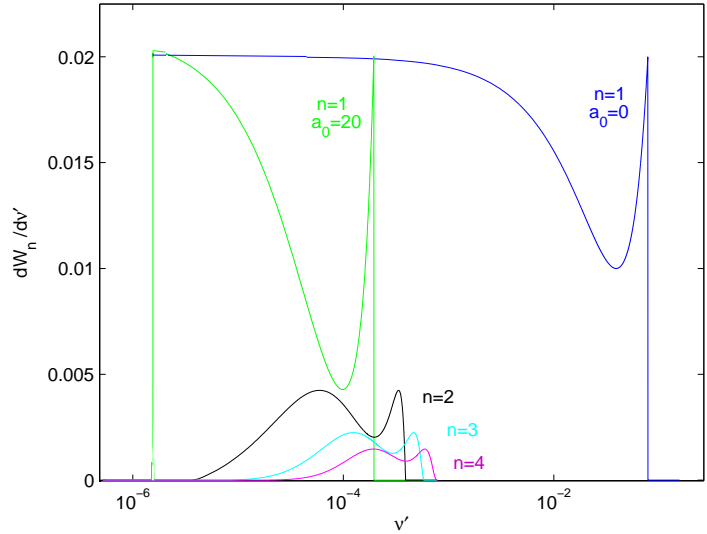}
\caption{\label{Fig:PartialRatesNU} Individual harmonic spectra ($n=1,\ldots 4$) for nonlinear Compton scattering at intensity $a_0 = 20$ compared to linear Compton scattering ($n=1$, $a_0=0$), as a function of $\nu'$. }
\end{center}
\end{figure}

\begin{figure}[!ht]
\begin{center}
\includegraphics[scale=0.45]{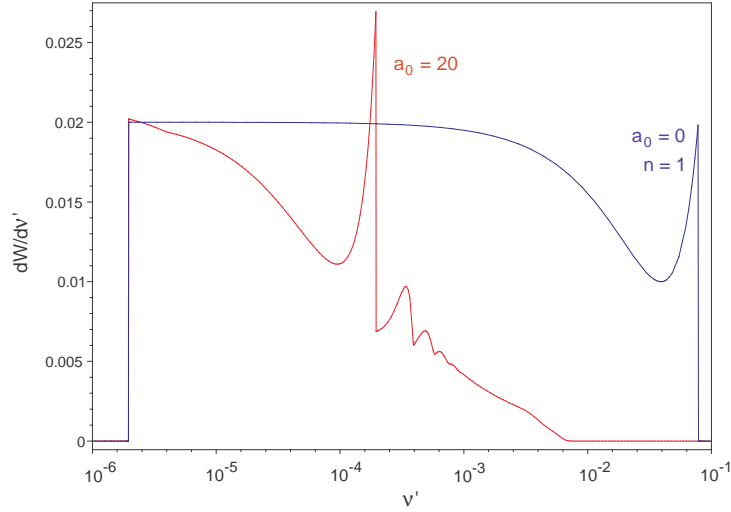}
\caption{\label{Fig:ratesumNU} Theoretical photon spectrum (sum of first 50 harmonics) for nonlinear Compton scattering at intensity $a_0 = 20$ compared to linear Compton scattering ($n=1$, $a_0=0$), as a function of $\nu'$. }
\end{center}
\end{figure}

Any actual Compton scattering experiment will be performed in a lab (frame) with the electrons either at rest (fixed target mode) or in motion. In what follows, we will assume the latter together with a head-on collision between laser pulse and electron beam (collider mode) as discussed in the previous section. In this case the kinematic invariants $x$ and $y_n$ from (\ref{X}) and (\ref{YN}) become functions of the scattered frequency $\nu'$ and the scattering angle $\theta$,
\bea
  x &=& \frac{(1 - \cos \theta) \nu'}{e^\zeta - (1 - \cos \theta)
  \nu'} \; , \label{XHO} \\
  y_n &=& \frac{2n\,\nu\, e^\zeta}{1 + a_0^2} \label{YNHO} \;
  .
\eea
Either the scattering angle $\theta$ or the frequency $\nu'_n$ may be eliminated via (\ref{NUPRIME}), allowing us to plot the emission rate as a function of $\nu'$ or $\theta$ respectively\footnote{The relationship between angle and frequency spectrum (\ref{NUPRIME}) is invertible \emph{provided} $\kappa_n \ne 0$. For $\kappa_n = 0$ the $n^\text{th}$ harmonic spectral range shrinks to a point (see below).}.  In this subsection we focus on the $\nu'$ dependence of the partial and total emission rates which are depicted in Fig.s~\ref{Fig:PartialRatesNU} and \ref{Fig:ratesumNU}, respectively. Similar plots (for $a_0$ of order one) have been obtained before in \cite{McDonald:1986zz,Tsai:1992ek,Bamber:1999zt,Ivanov:2004fi}.
Analytically the partial rates are 
\be
	\frac{\ud W_n}{\ud \nu'} = \frac{\ud W_n}{\ud x}\frac{\ud x}{\ud \nu'} = -\frac{\mathfrak{J}_n (z)}{\kappa_n}  \; .
\ee
The allowed range for $\nu'$ is given in (\ref{BS1}) and (\ref{INTERVALS}). The argument $z$ defined in (\ref{Z}) becomes a function of $\nu'$ via its dependence on
\be
  x \equiv x_n(\nu') = \frac{n\nu - \nu'}{\kappa_n - n\nu + \nu'} \; ,
\ee
upon eliminating $\theta$ from (\ref{XHO}) via (\ref{NUPRIME}).

For the parameters chosen ($\gamma = 10^2$, $\nu=2\times 10^{-6}$ and $a_0 = 20$) Fig.s \ref{Fig:PartialRatesNU} and \ref{Fig:ratesumNU} are fairly similar to their invariant pendants, Fig.s \ref{Fig:PartialRates20} and \ref{Fig:ratesum}. In particular, the previous shift in $x$ now corresponds to a red-shift of the linear Compton edge by a factor of $1 + a_0^2 \simeq 400$ from about 40 keV to 0.1 keV, i.e.\ from the hard to the soft X-ray regime. Note that the frequency range is still blue shifted relative to the incoming frequency $\nu$ (corresponding to the left-hand edge in Fig.s \ref{Fig:PartialRatesNU} and \ref{Fig:ratesumNU} given by $\nu = 2 \times 10^{-6}$). Again, there is a noticeable enhancement of the total emission rate at $\nu_n^\prime (\pi) \simeq 4 \gamma^2 \nu/a_0^2$, cf.\ (\ref{NUPRIME-HE}), due to the generation of peaks corresponding to higher harmonics, $n > 1$, with the peak values decreasing rapidly with $n$. We note that the edge values of the higher harmonics which are clearly visible in Fig.~\ref{Fig:PartialRatesNU} get washed out by the superposition of more and more partial rates $\ud W_n$ in Fig.~\ref{Fig:ratesumNU}. This will reduce the visibility of the associated maxima, as will, of course, all sorts of background effects which have not been included in the theoretical analysis above.

\begin{figure}[!ht]
\begin{center}
\includegraphics[scale=0.8]{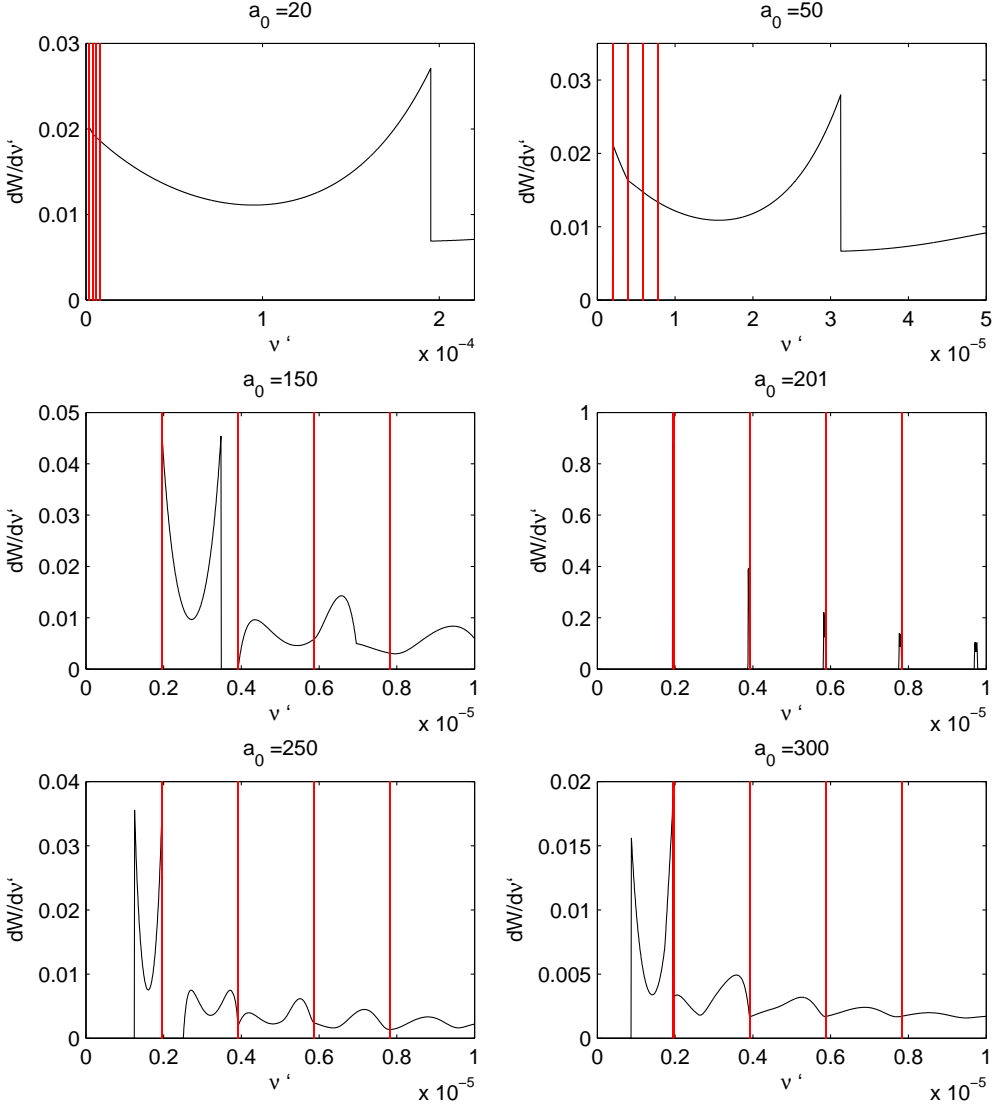}
\caption{\label{Fig:MOVIE} (Color online) Theoretical photon spectra for nonlinear Compton scattering for different values of $a_0$ ($\gamma = 100$) and incoming frequency $\nu = 2 \times 10^{-6}$. The vertical (red) lines correspond to frequencies $n\nu$.}
\end{center}
\end{figure}

The properties of the photon spectrum depend crucially on electron parameters ($\beta$, $\gamma$ or $\zeta$) characterising the lab frame and, in particular, the intensity parameter $a_0$. To illustrate this dependence along with the discussion of Subsection \ref{subsect:KINEMATICS} we have calculated the photon spectra as a function of $a_0$, ranging from $a_0 = 20$ up to $300$. The outcome is depicted in the movie-like sequence of plots of Fig.~\ref{Fig:MOVIE}. As $\gamma = 100$ the critical $a_0$ from (\ref{a0-crit}) defining the CM frame of the first harmonic is $a_{0\mathrm{crit}}(1) \simeq 200$ corresponding to the fourth plot in Fig.~\ref{Fig:MOVIE}. There, the lower harmonic spectrum collapses to lines located at the individual harmonics with frequencies $\nu_n' = n\nu$ (marked by red vertical lines throughout).

If we go through the whole sequence the following picture emerges. For small $a_0 < a_{0\mathrm{crit}}(n)$, all harmonic ranges with counting label less than $n$ are blue shifted. Plots 1 and 2 show the harmonic range for $n=1$ (and part of $n=2$), both to the right of their red end edges ($\nu$ and $2\nu$, respectively). The right-hand, blue end, maximum of the fundamental range is enhanced due to contributions of higher harmonics. For $a_0$ approaching its critical value the harmonic ranges shrink, and a gap between the first and second appears (Plot 3) so that the fundamental maxima become of equal height. At $ a_0 = a_{0\mathrm{crit}} (1) \simeq 200$ the first harmonic range shrinks (almost) to a point, with the neighbouring ranges also becoming very narrow (Plot 4). Once $a_0 (1)$ becomes super-critical, all harmonic ranges are red-shifted (i.e.\ located to the left of the vertical (red) lines, $\nu'_n < n\nu$), with the ranges increasing again and gaps closing (Plots 5 and 6). In Plot 6, the first and second harmonics overlap again, leading to maxima of different height, with the one at $\nu_1' = \nu$ being the larger.

Thus, by tuning $a_0$ we effectively change frames of references with $a_{0\mathrm{crit}} (1)$ representing the border between inverse Compton scattering (blue-shift) and Compton scattering (red-shift).

\subsection{Lab kinematics: angular dependence}

As mentioned earlier, the emission rates may be considered as functions of either scattered frequency $\nu_n'$, or scattering angle $\theta$,  the two being related via (\ref{NUPRIME}). In terms of the scattering angle $\theta$ the rates become
\be \label{RATE-THETA}
  \frac{\ud W_n}{\ud \Omega} = \frac{\ud W_n}{\ud x}\frac{\ud x}{\ud \Omega} = \frac{e^\zeta}{n \nu (1 - \cos \theta)^2} \frac{x_n^2}{(1 + x_n)^2} \, \mathfrak{J}_n (z_n) \; , \quad 0 < \theta < \pi \; ,
\ee
where $x_n \equiv x$ (for the $n$-th harmonic) and $z_n$ are to be viewed as functions of $\theta$ (see below). Our angular measure is $\ud\Omega \equiv \ud\theta \sin \theta $, which is the solid angle measure up to a factor of $2\pi$, as the azimuthal angle $\phi$ does not contribute due to axial symmetry. Note that this is different for linear polarisation or, more generally, if there is another preferred direction which, for instance, could be induced by noncommutative geometry \cite{Ilderton:2009}.

\begin{figure}[!ht]
\begin{center}
\includegraphics[scale=0.6]{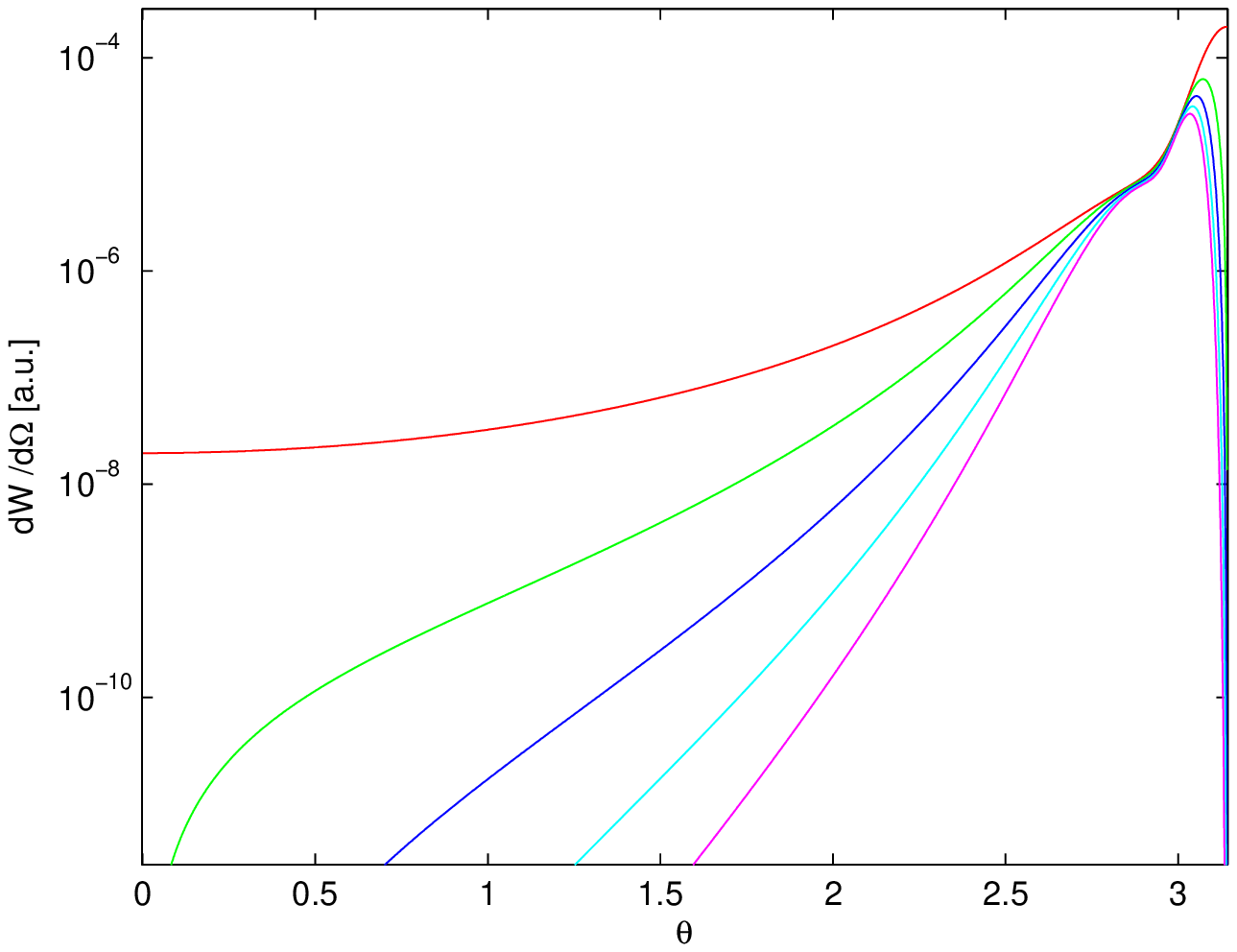} \includegraphics[scale=0.6]{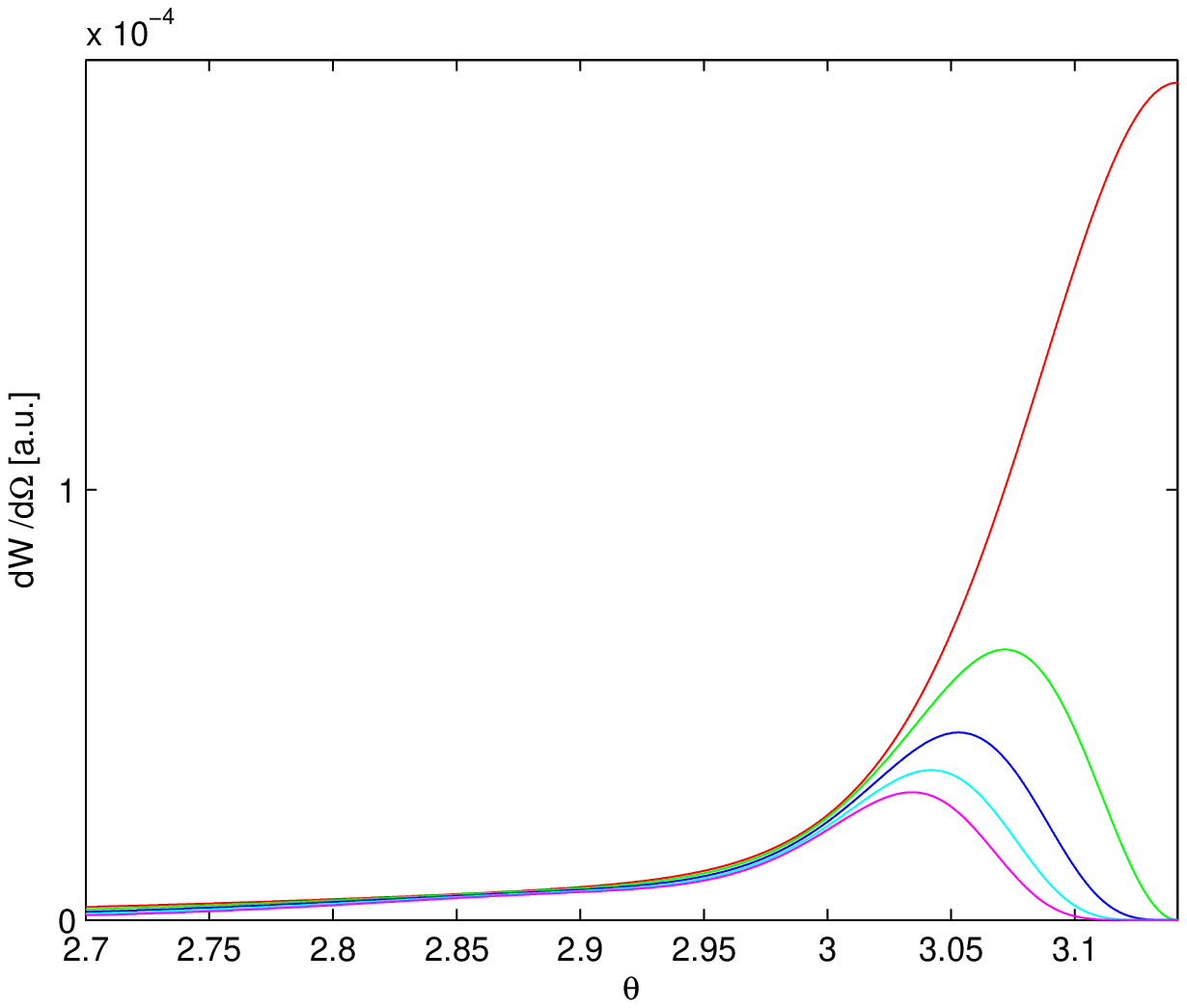}
\caption{\label{Fig:HARMONICS-THETA} Theoretical photon spectrum for the first five individual harmonics as a function of scattering angle $\theta$. Parameters: $\gamma = 100$, $a_0 = 20$; LEFT: vertical scale logarithmic; RIGHT: vertical scale linear, zoomed into range $2.7 < \theta < \pi$.}
\end{center}
\end{figure}

In terms of their angular dependence the various invariants may all be expressed, using (\ref{NUPRIME}) and (\ref{X}), in terms of the variable $x_1$ defined by
\be
  x_n (\theta) \equiv n x_1 (\theta) = \frac{2n\nu (1 - \cos \theta)}{e^\zeta (1 + \cos \theta) + e^{-\zeta}(1 + a_0^2) (1 - \cos \theta)} \; ,
\ee
with $x_1$ between $x_1(0) = 0$ and $x_1(\pi)=y_1$ as in (\ref{YN}), where
\be
  y_1 = \frac{2 \nu e^\zeta}{1 + a_0^2} \; .
\ee
The argument of $\mathfrak{J}_n $ in (\ref{RATE-THETA}) becomes
\be
  z_n (\theta) \equiv n z_1 (\theta) = 2 n \frac{a_0}{\sqrt{1 + a_0^2}} \, \sqrt{r (1-r)}  \; ,
\ee
where we have introduced the rescaled variable
\be
  r \equiv x_1/y_1 = \frac{e^{-\zeta}(1 + a_0^2) (1 - \cos \theta)}{e^\zeta (1 + \cos \theta) + e^{-\zeta}(1 + a_0^2) (1 - \cos \theta)}  \; , \quad 0 \le r \le 1 \; .
\ee
As a result, $z_1$ becomes maximal for $r = 1/2$ and so $z_1$ is less than unity,
\be \label{Z1MAX}
  z_1 \le \frac{a_0}{\sqrt{1 + a_0^2}} < 1 \; ,
\ee
which will be important later when we discuss the convergence of the emission rate sum. Solving $r(\theta_0) = 1/2$ we find $z_1$ is maximised at the angle
\be \label{CONEANGLE}
  \theta_0 = \arccos \frac{1 + a_0^2 - e^{2\zeta}}{ 1 + a_0^2 + e^{2\zeta}} \;.
\ee
We will now relate these results to the emission spectra as functions of $\theta$. In Fig.~\ref{Fig:HARMONICS-THETA} we show the angular distribution of the photon yield, as determined by (\ref{RATE-THETA}), for the lowest individual harmonics, $n=1, \ldots, 5$. For the parameters chosen ($\gamma = 10^2$, $\nu=2\times 10^{-6}$ and $a_0 = 20$) the largest signal is due to the fundamental harmonic, $n=1$. This is also the only one contributing on axis, i.e.\ in the forward and backward directions, $\theta = 0$ and $\pi$, respectively. For the classical intensity distribution this was also found by Sarachik and Schappert \cite{Sarachik:1970ap}. Thus, in particular, real backscattering at $\theta = \pi$ only occurs for $n=1$, while for the higher harmonics one has `dead cones' with an opening angle of about 0.1 radians, slightly increasing with harmonic number $n$, as seen from the magnified plot in Fig.~\ref{Fig:HARMONICS-THETA} (right panel).

The dead cones are controlled by the angle $\theta_0$ from (\ref{CONEANGLE}): their opening angles are bounded by $\theta_0' \equiv \theta_0 - \pi$. For $1 \ll a_0^2 \ll \gamma^2$ the former are quite narrow such that most of the radiation (in particular the location of the maxima at $\theta_0$) is near backward
\footnote{We mention in passing that the situation for linear polarisation is different. As pointed out by Esarey et al. \cite{Esarey:1993zz} for Thomson scattering with linearly polarised photons, \textit{odd} harmonics do get backscattered (no dead cones).}. Quantitatively one finds that the dead cone opening angles are less than
\be \label{THETAPRIME}
  \theta_0'  \simeq a_0/\gamma \ll 1 \; ,
\ee
which, for the parameters of Fig.~\ref{Fig:HARMONICS-THETA} corresponds to $\theta_0' \simeq 0.2$ radians. (For the intensity distribution of classical radiation the relation (\ref{THETAPRIME}) was found in \cite{Esarey:1993zz}.)

To determine the total emission rate we have to sum (\ref{RATE-THETA}) over all harmonic numbers, $n$. It is not entirely obvious that the ensuing series converges. To prove this we employ the Bessel function identity \cite{Abramowitz:1972},
\be
  J_{n \pm 1}(z) = \frac{n}{z} J_n (z) \mp J_n^\prime (z) \; ,
\ee
the prime denoting the derivative with respect to the argument $z$, in order to rewrite $\mathfrak{J}$ in terms of $J_n^2$ and $J_n^{\prime 2}$,
\be
  \mathfrak{J}(z_n) = 2 J_n^2 (nz_1) \, \left[ - \frac{2}{a_0^2} + \left( 2 + \frac{n^2 x_1^2}{1 + nx_1} \right) \left( \frac{1}{z_1^2} - 1 \right) \right] + 2 J_n^{\prime 2} (nz_1) \,  \left( 2 + \frac{n^2 x_1^2}{1 + nx_1} \right) \; .
\ee
According to (\ref{RATE-THETA}), in the rates this is multiplied with an $n$-dependent factor $n/(1 + nx_1)^2$. Thus, upon summation, we encounter series of the form
\be
	\sum_{n > 0} \frac{n^N}{(1 + nx_1)^M} \, J_n^2 (nz_1) \quad\text{and}\quad \sum_{n > 0} \frac{n^N}{(1 + nx_1)^M} \, {J'}_n^2 (nz_1)\;,
\ee
where $N \in \{1,3\}$ and $M \in \{2,3\}$. We can easily bound these series from above, for example
\bea
  \sum_{n > 0} \frac{n}{(1 + nx_1)^2} \, J_n^2 (nz_1) &<&  \sum_{n > 0} n J_n^2 (nz_1) \equiv S_1 \; , \label{KAPTEYN1} \\
  \sum_{n > 0} \frac{n^3}{(1 + nx_1)^3} \, J_n^2 (nz_1) &<& \sum_{n > 0} n^3 J_n^2 (nz_1) \equiv S_3 \; , \label{KAPTEYN2}
\eea
(and likewise for $J_n^{\prime 2}$). The series $S_1$ and $S_3$ on the right hand side are examples of Kapteyn series \cite{Kapteyn:1893} which are known to converge. Remarkably, some also have analytic expressions for the sum. These results do not seem particularly common, so we collect them in an appendix. Although we have not yet been able to explicitly perform our sums (which have a more complicated $n$--dependence than the Kapteyn series) we can now be confident that they converge. This is an extremely satisfying result confirming the validity of the background field picture we have employed and our analysis based around the summation of individual harmonics.

\begin{figure}[!ht]
\begin{center}
\centering\includegraphics[scale=0.6]{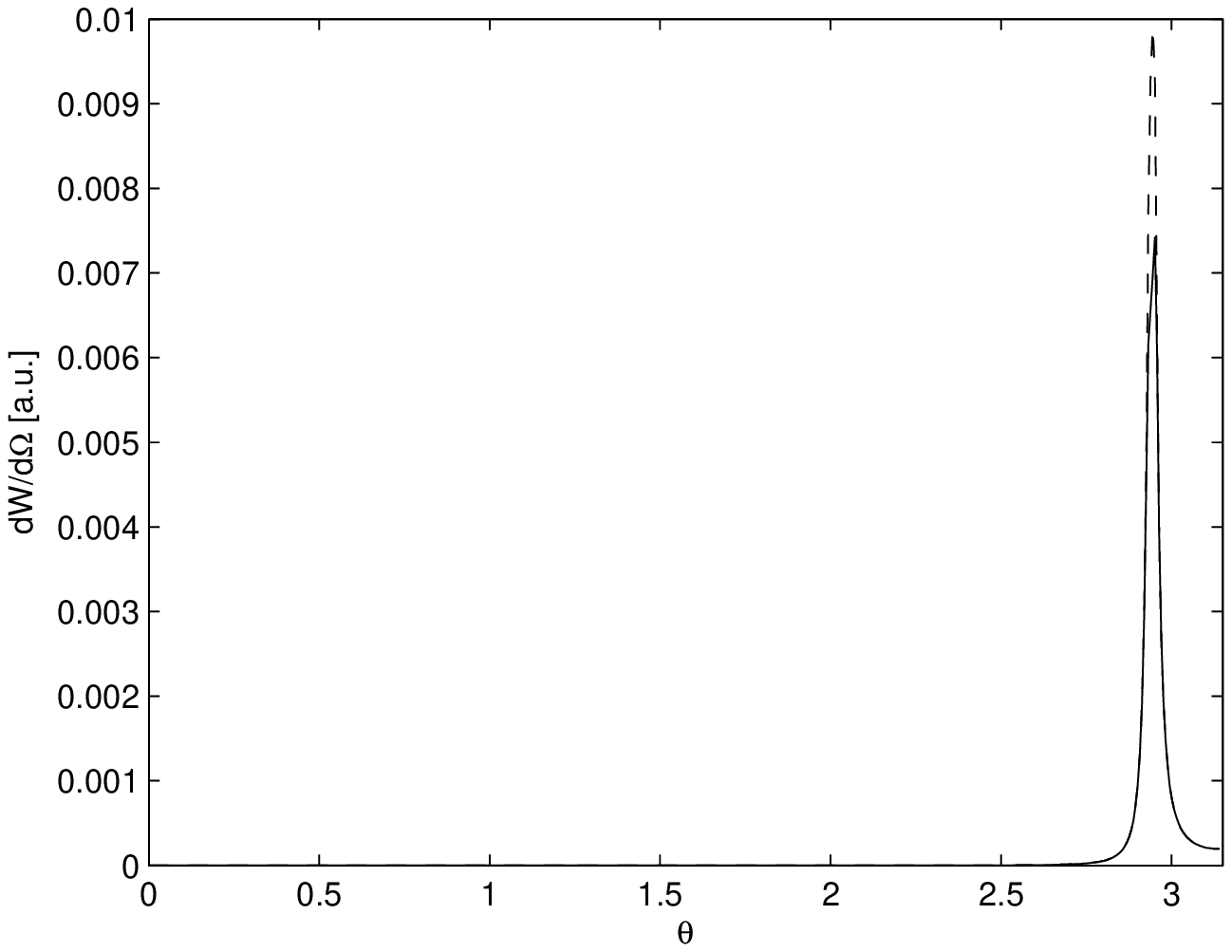}\includegraphics[scale=0.6]{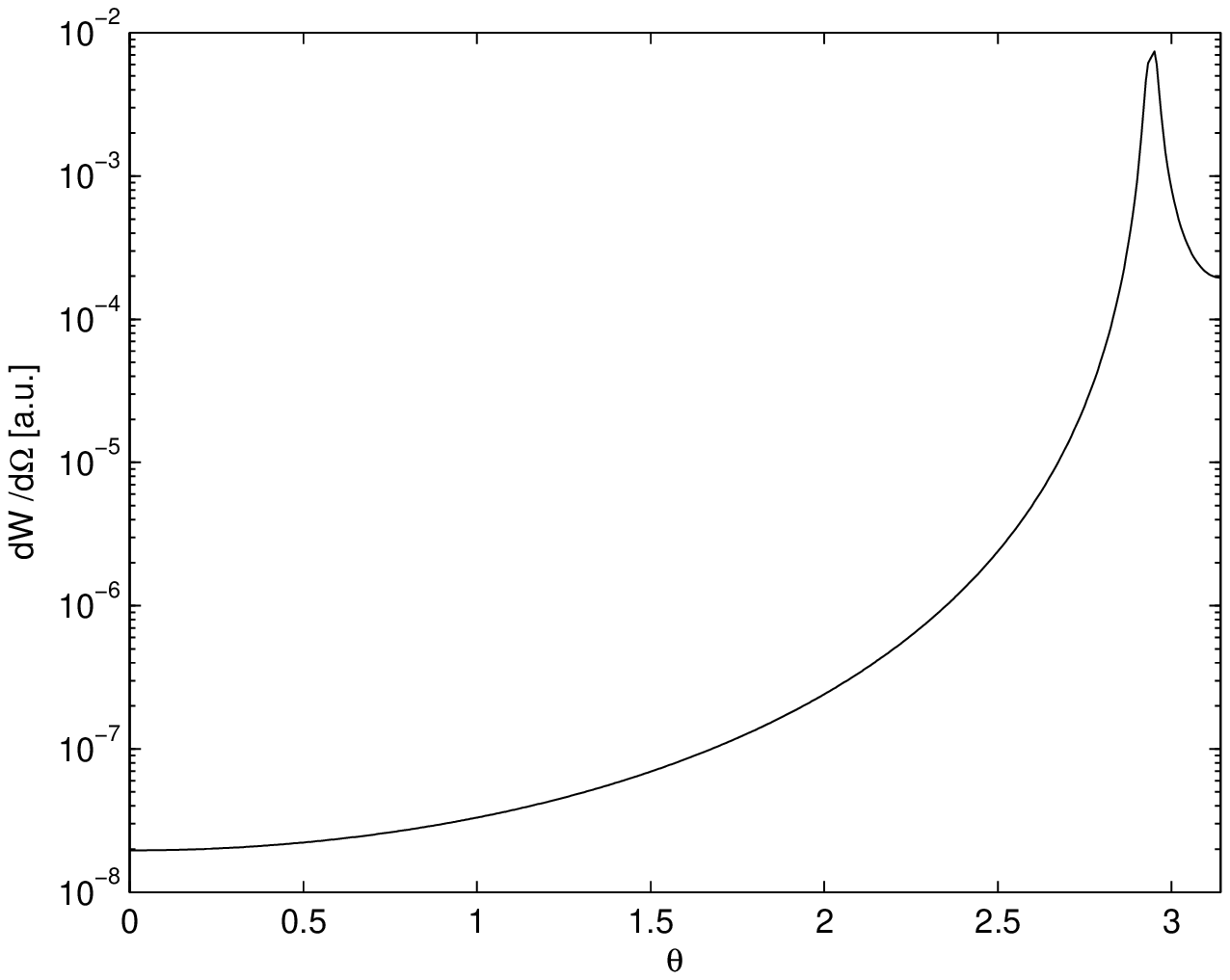}
\caption{\label{Fig:LIN-SUM-THETA} Theoretical photon spectrum. Parameters: $\gamma = 100$, $a_0 = 20$. LEFT: vertical scale linear, harmonics summed up to $n=5000$ (full line) and $n=10000$ (dashed line). RIGHT: vertical scale logarithmic, harmonics summed up to $n=5000$.}
\end{center}
\end{figure}

Lerche and Tautz in \cite{Lerche:2008} state that a summation of the first 1000 terms in Kapteyn series like (\ref{KAPTEYN1}) or (\ref{KAPTEYN2}) yields errors below $10^{-6}$ for $z_1 \lesssim 0.95$. We need to include $z_1$ values closer to one where the convergence rate is at its lowest. This occurs near the angle $\theta_0$ defined in (\ref{CONEANGLE}). Increasing the maximum harmonic number from 5000 to 10000 yields basically identical plots except that the height of the narrow peak at $\theta_0$ increases as shown in Fig.~\ref{Fig:LIN-SUM-THETA} (left panel). The maximum is indeed located at $\theta = \theta_0 = 2.94$ (or  $\theta_0' \simeq a_0/\gamma \simeq 0.2$) as given in (\ref{CONEANGLE}) and (\ref{THETAPRIME}). The shoulder near $\theta = \pi$ ($\theta' = 0$) is entirely due to the fundamental harmonic ($n=1$).

\begin{figure}[!ht]
\begin{center}
\includegraphics[scale=0.7]{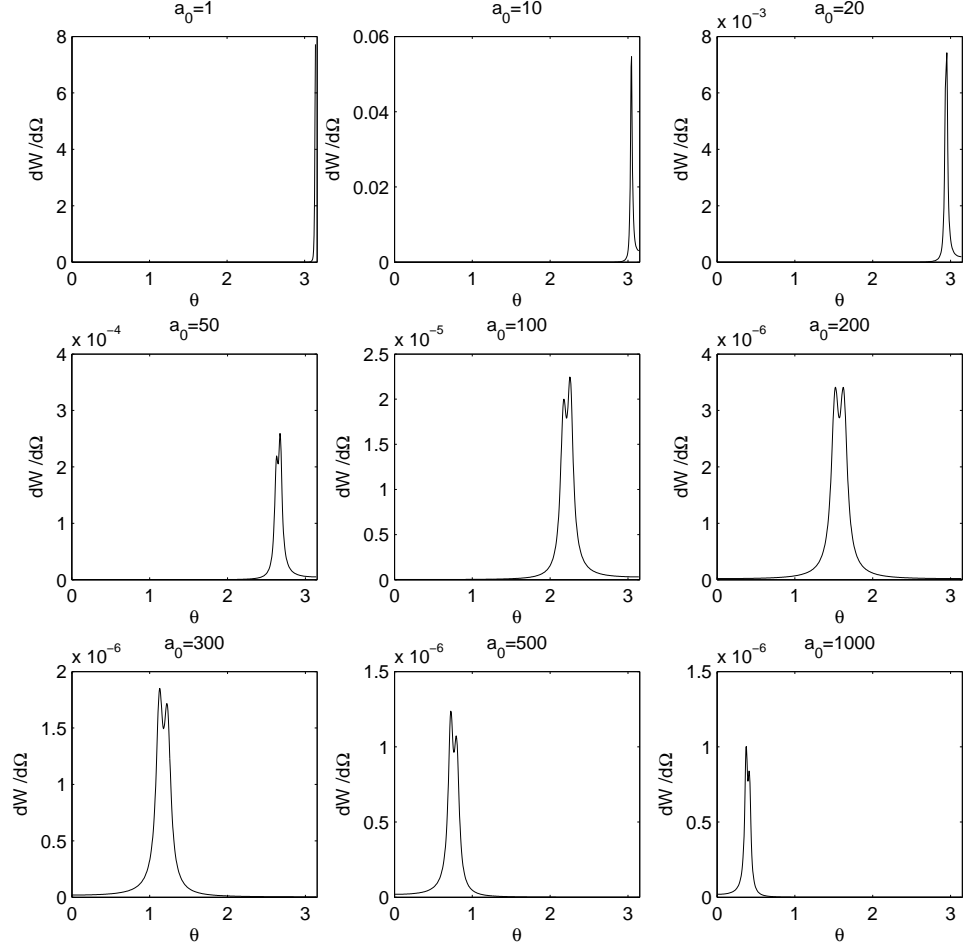}
\caption{\label{Fig:A0-THETA} Theoretical photon spectrum as a function of $\theta$, harmonics summed up to $n=5000$ for different values of $a_0$ ($\gamma = 100$); vertical scale linear.}
\end{center}
\end{figure}
\begin{figure}[!ht]
\begin{center}
\includegraphics[scale=0.7]{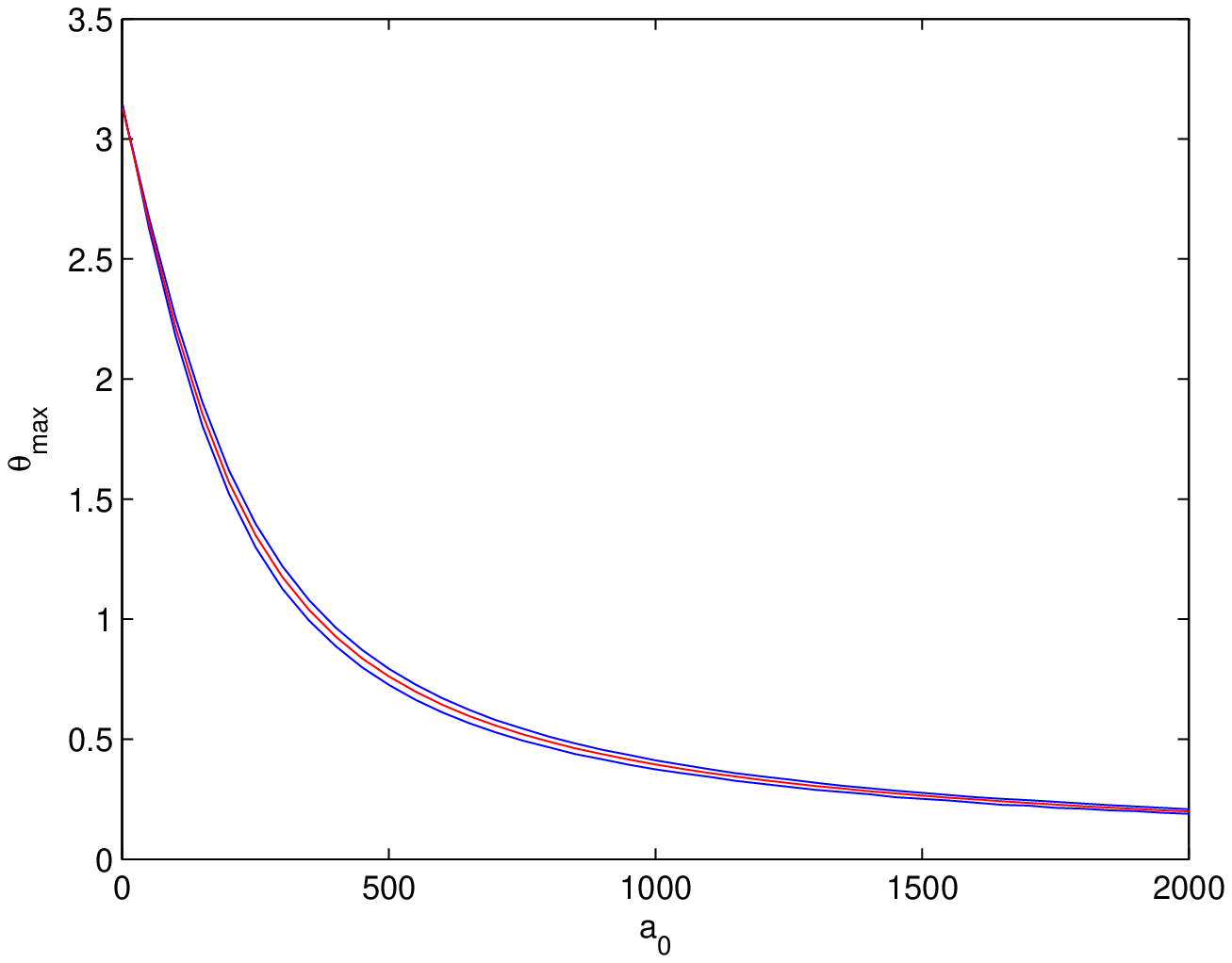}
\caption{\label{Fig:PEAKPOS} The angular position of the maximum emission rates as a function of intensity, $a_0 \ge 1$ (blue/upper and lower curves), and the angle $\theta_0$ which defines the maximum value of $z_1$ (red/middle curve). Harmonics summed up to $n=5000$, $\gamma = 100$.}
\end{center}
\end{figure}

Finally, we again vary $a_0$ and plot a movie of the angular distribution for fixed $\gamma = 100$ in Fig.~\ref{Fig:A0-THETA}. The main features are (i) a propagation of the main peak from near backward direction (when $a_0 \ll 2 \gamma$) to near forward direction (when $a_0 \gg 2\gamma$) consistent with the formula (\ref{CONEANGLE}) for $\theta_0$, (ii) the appearance of a double peak which (iii) becomes symmetric for $a_0 \simeq 2\gamma$ at an angle $\theta_0 = \pi/2$. The latter situation corresponds to $\cos \theta_0 = 0$, hence
\be
  a_0^2 = e^{2\zeta} - 1 \simeq e^{2 \zeta} \simeq 4 \gamma^2 \; , \quad (1 \ll a_0^2 \ll \gamma^2) \; .
\ee
This latter value (approximately) coincides with the critical $a_0$ of (\ref{a0-crit-approx}). The locations of the two peaks in the spectrum are plotted in Fig.~\ref{Fig:PEAKPOS}, along with the angle $\theta_0$ given in (\ref{CONEANGLE}), as a function of $a_0$. It is clear from this plot that the maximum value of $z_1$ corresponds to the local minimum between the two peaks.

\subsection{Thomson limit: emission rate and intensity}

\begin{figure}[!ht]
\begin{center}
\includegraphics[scale=0.7]{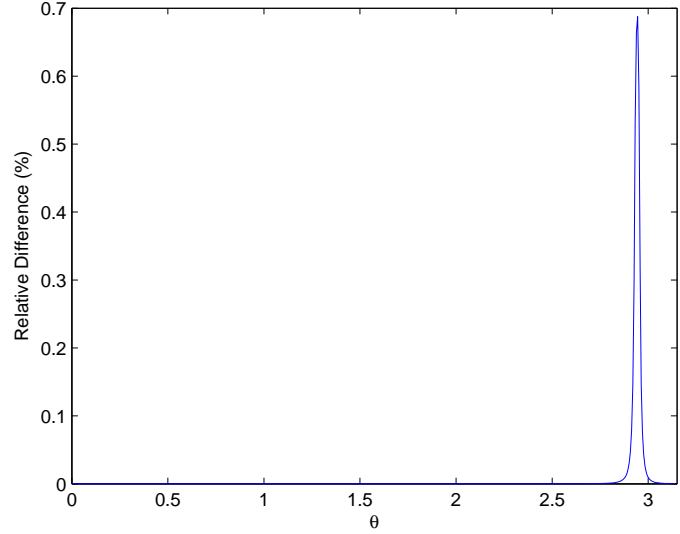}
\caption{\label{Fig:THOMSONCOMPTON} Relative difference of photon emission rates $|$Compton $-$ Thomson$|$/Compton as a function of scattering angle $\theta$. Harmonics summed up to $n=10000$, $\gamma = 100$, $a_0 = 20$.}
\end{center}
\end{figure}

At this point one should mention that thorough discussions of the intensity distributions employing \textit{classical} radiation theory have appeared before \cite{Sarachik:1970ap,Esarey:1993zz}. It is useful to check that our quantum calculations based on the Feynman diagrams of Fig.~\ref{Fig:NLC} describing nonlinear Compton scattering reproduce the results for nonlinear \textit{Thomson} scattering in the classical limit. According to Nikishov and Ritus \cite{Nikishov:1963} the classical limit is given by
\be \label{CLASSICAL}
  y_n = \frac{2n p \cdot k}{m_*^2} \ll 1 \; ,
\ee
which is just the statement that $m_*$ is the dominant energy scale. Note that this can be achieved by having large $a_0$ and may be counterbalanced by large $n$. Hence, harmonics with sufficiently large harmonic number $n$ will behave non-classically (if they are observable at all despite their suppression). As $y_n$ is the upper bound for $x_n$ (\ref{CLASSICAL}) may equivalently be formulated as
\be \label{CLASSICALLIMIT}
  x_n  \ll 1 \; ,
\ee
such that we may neglect $x_n = nx_1$ on the left hand sides of (\ref{KAPTEYN1}) and (\ref{KAPTEYN2}) which hence \textit{coincide} with $S_1$ and $S_3$ in the classical limit. Even if (\ref{CLASSICALLIMIT}) no longer holds (i.e.\ for large $n$) contributions to the sum are still suppressed by $J_n^2$. Comparing the quantum and classical (Compton vs.\ Thomson) rates by evaluating all sums numerically the graphs are indistinguishable. Plotting the relative difference for our parameter values one finds a small discrepancy near $\theta = \theta_0$, of the order of $1\%$ (see Fig.~\ref{Fig:THOMSONCOMPTON}). Note that the classical series $S_1$ and $S_3$ have a slightly slower rate of convergence (in particular near $z_1 = 1$, i.e.\ $\theta = \theta_0$) where the suppression is mainly provided by $J_n^2 (nz_1)$, hence least efficient at $z_1 = 1$. We have found for instance, that the peak in Fig.~\ref{Fig:THOMSONCOMPTON} increases from $0.4 \%$ to $0.7\%$ when we increase the maximum $n$ from 5000 to 10000. Nevertheless, Fig.~\ref{Fig:THOMSONCOMPTON} provides a nice confirmation that for high intensity optical lasers the background can indeed be treated as classical to a very good approximation.

We are left with relating photon production probabilities $\ud W_n$ to intensities $\ud I_n$. This problem has also been addressed by Nikishov and Ritus \cite{Nikishov:1963} who state that the intensity is given by the zero component of the radiation 4-vector,
\be
  P_\mu \equiv \sum_{n>0} \int \ud W_n \, k_\mu^\prime \; .
\ee
We thus have $\ud I_n = m \nu' \ud W_n$ or
\be\label{DIN}
  \frac{\ud I_n}{d\theta} = m e^{2\zeta} \, \frac{\nu^2}{\sin^2 \theta} \, \frac{z_1^3}{a_0^3} \, \frac{n^2}{(1 + nx_1)^3} \, \mathfrak{J}_n (nz_1) \; .
\ee
Compared to (\ref{RATE-THETA}) we thus have an additional factor $n/(1 + nx_1)$. In the classical limit, $nx_1 \ll 1$, this is just $n$ so that (\ref{DIN}) is bounded not by the Kapteyn series $S_1$ and $S_3$, but by the analytically known series $S_2$ and $S_4$ as given in the appendix.

\section{Conclusions}\label{Concs}

In this paper we have (re)assessed the prospects for observing intensity effects in Compton scattering. The physical scenario assumed is the collision of a high-intensity laser beam with an electron beam of sufficiently high energy ($\gamma \gtrsim 10^2$) produced in a conventional accelerator or by a suitable laser plasma acceleration mechanism. In technical terms we were interested in the features present in cross sections or photon emission rates which are enhanced with increasing dimensionless laser amplitude, $a_0 = ea/m$ where $a$ is the magnitude of the laser vector potential. The possible effects are of a mostly classical nature, being fundamentally due to the mass shift, $m^2 \to m_*^2 = m^2 (1 + a_0^2)$ caused by the relativistic quiver motion of an electron in a laser field. Ranked in order of their relevance the main intensity effects are: (i) a red-shift of the kinematic Compton edge for the fundamental harmonic, $\omega' = 4 \gamma^2 \omega \to 4 \gamma^2 \omega/a_0^2$ for the parameters we have used, (ii) the appearance of higher harmonic peaks ($n>1$) in the photon spectra and (iii) a possible transition from inverse Compton scattering ($\omega' > \omega$) to Compton scattering ($\omega' < \omega$) upon tuning $a_0$. The red-shift (i) may be explained in terms of the larger effective electron mass, $m_* > m$, the generation of which costs energy that is missing when it comes to `boosting' the photons to higher frequencies. This has, for instance, an impact on X-ray generation via Compton backscattering. To avoid significant energy losses (reducing the X-ray frequency) the amplitude $a_0$ should probably not exceed unity significantly. However, one is certainly dealing with a fine-tuning problem here, as item (ii), the generation of higher harmonics, improves the X-ray beam energy distribution. For $a_0 > 1$ there is a larger photon yield due to superposition of the harmonics and the full width at half maximum goes down. As a result, the X-rays tend to become more monochromatic once higher harmonics become involved. Item (iii), the transition from inverse to ordinary Compton scattering, once $a_0$ increases beyond $2\gamma$ illustrates the energy `loss' just mentioned. When $a_0 \simeq 2 \gamma$ the lab frame can be interpreted as an intensity dependent centre-of-mass frame for which $\omega_n' = n\omega$, at least for low harmonics. Thus there is no longer an energy gain of the emitted photons: the laser beam has become so `stiff' that, in this frame, electrons begin to bounce back from it (gaining energy) rather than vice versa.

The next step is to actually perform the experiments required for measuring the effects listed above. We emphasise that nonlinear Compton scattering provides a unique testing ground for strong-field QED as the process is not suppressed in terms of $\alpha$ or $E/E_c$, by powers or exponentially. Hence, the experiments at Daresbury ($\gamma \simeq 50$, $a_0 \simeq 2$) \cite{Priebe:2008} and the FZD ($\gamma \simeq 80$, $a_0 \simeq 20$) planned for the near future should indeed be able to see the effects analysed in this paper. This will provide crucial evidence for the validity of the approach to strong-field QED adopted here, based on the electron mass shift, the Volkov solution and the Furry picture.

\subsubsection*{Acknowledgements}

The authors thank  F.~Amiranoff, M.~Downer, G.~Dunne, H.~Gies, B.~K{\"a}mpfer, K.~Langfeld, M.~Lavelle, K.~Ledingham, M.~Marklund, D.~McMullan, G.~Priebe, R.~Sauerbrey, G.~Schramm, D.~Seipt, V.~Serbo and A.~Wipf for discussions on various aspects of strong-field QED. A.I.\ is supported by an IRCSET postdoctoral fellowship, C.H.\ by an EPSRC doctoral training award (Ref EP/P502675/1). A.I.\ thanks the Plymouth Particle Theory Group for hospitality.

\appendix

\section{Kapteyn Series}
The Kapteyn series \cite{Kapteyn:1893} (see also \cite[Ch.~XVII]{Watson:1922}) of the second kind involve squares of Bessel functions or their derivatives. We use the notation
\bea\label{KAPGEN}
	S_N &\equiv \sum\limits_{n>0} n^N J_n^2(n z_1)\;, \\
	S'_N &\equiv \sum\limits_{n>0} n^N {J'_n}^2(n z_1)\;, \\
\eea
where $0<z_1<1$ in keeping with our earlier discussion. The sums with a closed form expression are
\bea
  S_{-2} &\equiv& \sum_{n > 0} n^{-2} J_n^2 (nz_1) = \frac{z_1^2}{4} \; , \\
  S_0 &\equiv& \sum_{n > 0} J_n^2 (nz_1) = \frac{1}{2\sqrt{1-z_1^2}}-\frac{1}{2}\; , \\
  S_2 &\equiv& \sum_{n > 0} n^2 J_n^2 (nz_1) = \frac{z_1^2(4 + z_1^2)}{16(1-z_1^2)^{7/2}} \; , \\
  S_4 &\equiv& \sum_{n > 0} n^4 J_n^2 (nz_1) = \frac{z_1^2(64 + 592 z_1^2 + 472 z_1^4 + 27 z_1^6)}{256 (1 - z_1^2)^{13/2}} \; .
\eea
The first is a result of Nielsen \cite{Nielsen:1904} according to Schott who derived the second and third results \cite[p.122]{Schott:1912}, while the fourth can be found in \cite{Lerche:2007} (note that our notation differs from that paper, which also contains a typographical error in their equation (24) for $S_2$). The sums involving $J_n'$ are
\bea
  S_2^\prime &\equiv& \sum_{n > 0} n^2 J_n^{\prime 2} (nz_1) = \frac{4 + 3z_1^2}{16(1-z_1^2)^{5/2}} \; , \\
  S_4^\prime &\equiv& \sum_{n > 0} n^4 J_n^{\prime 2} (nz_1) = \frac{64 + 624 z_1^2 + 632 z_1^4 + 45 z_1^6}{256 (1 - z_1^2)^{11/2}} \; ,
\eea
given in \cite{Sarachik:1970ap} and \cite{Lerche:2007}, respectively. The latter paper also gives a double integral representation for the series $S_{-1}$ (there denoted $F_+$).
Referring to a theorem by Watson \cite{Watson:1917} the authors of \cite{Lerche:2008} derive an iterative scheme for higher-order Kapteyn series, giving, for example,
\bea
  S_1 &=& \frac{1}{1 - z_1^2} \left( z_1 \pad{}{z_1} \right)^2 S_{-1} \; , \\
  S_3 &=& \frac{1}{1 - z_1^2} \left( z_1 \pad{}{z_1} \right)^2 S_{1} \; .
\eea
%
%


\providecommand{\href}[2]{#2}\begingroup\raggedright\endgroup

\end{document}